\documentclass[apa,pre,twocolumn,showkeys, floatfix]{revtex4-1}
\usepackage[utf8]{inputenc}
\usepackage{amsmath}
\usepackage{graphicx}
\usepackage{subcaption}
\usepackage{amsfonts}
\usepackage{amssymb}
\usepackage{epstopdf}
\usepackage{xcolor}

\begin{document}

\title{Thermodynamic efficiency of contagions: A statistical mechanical analysis of the SIS epidemic model}
\author{Nathan Harding$^1$}
\email[Correspondence should be sent to: ]{nathan.harding@sydney.edu.au}
\author{Ramil Nigmatullin$^1$}
\author{Mikhail Prokopenko$^{1,2}$}
\affiliation{$^1$Centre for Complex Systems, Faculty of Engineering and IT\\ University of Sydney, Sydney, NSW 2006, Australia\\
$^2$Marie Bashir Institute for Infectious Diseases and Biosecurity\\ University of Sydney, Westmead, NSW 2145, Australia}

\begin{abstract}
We present a novel approach to the study of epidemics on networks as thermodynamic phenomena, quantifying the thermodynamic efficiency of contagions, considered as distributed computational processes. Modelling SIS dynamics on a contact network statistical-mechanically, we follow the Maximum Entropy principle to obtain steady state distributions and derive, under certain assumptions, relevant thermodynamic quantities both analytically and numerically. In particular, we obtain closed form solutions for some cases, while interpreting key epidemic variables, such as the reproductive ratio $R_0$ of a SIS model, in a statistical mechanical setting. On the other hand, we consider configuration and free entropy, as well as the Fisher Information, in the epidemiological context. This allowed us to  identify criticality and distinct phases of epidemic processes.
For each of the considered thermodynamic quantities, we compare the  analytical solutions informed by the Maximum Entropy principle with the numerical estimates for SIS epidemics simulated on Watts-Strogatz random graphs.  
\end{abstract}
\maketitle

\section{Introduction}
Various real-world crises and disruptive events, such as epidemics, cascading technological failures, ecological and economic tipping points, can be quantitatively studied as critical phenomena, so that the corresponding critical thresholds can be identified, predicted and used in planning suitable crisis interventions (e.g., vaccinations and quarantine, power-grid safety margins, climate change policies). Modelling critical dynamics typically involves analysis of sensitivities to initial conditions and the overall spatiotemporal behaviour at the system level. Within physics, such behaviour is characterised in terms of the control and order parameters, allowing the modellers to investigate phase transitions. A canonical example is a second-order phase transition in a ferromagnetic system, which separates two qualitatively different phases: a disordered paramagnetic phase characterised by the
absence of net magnetisation in the high-temperature
regime, and the ordered ferromagnetic phase with a
net magnetisation in the low-temperature regime. Importantly, the change between these phases is sudden and is driven by varying the control parameter (temperature). The resulting magnetisation outcome is traced by the order parameter: the net magnetisation vector which quantifies the emerged preferred direction in space. Formally, a phase transition manifests itself as ``a sharp change in the properties (state) of a substance (system)'' occurring when ``there is a singularity in the free energy or one of its derivatives''~\cite{pastor2015epidemic}.

Like many other fields of research, epidemiology has also been drawing on the results obtained within statistical physics in terms of critical thresholds and phase transitions.  For example, several studies have successfully modelled epidemic spread as a specific example of percolation in networks~\cite{newman1999scaling, moore2000epidemics, newman2002spread, sander2002percolation, meyers2007contact}. Under certain (fairly strong) assumptions, the problem of when an epidemic takes place becomes equivalent to a standard percolation problem on a graph, whose objective is to compute the fraction of sites that must be occupied before the formation of a ``giant component'' of connected sites. The size of such giant component scales extensively with the total number of sites~\cite{moore2000epidemics}, demonstrating scale-invariance, a well-known feature observed during critical regimes.
 
A critical threshold that is often studied in epidemiology is the epidemic threshold defined with respect to the pathogen's reproductive ratio ($R_0$), that is, the number of secondary infections generated on average, within a susceptible population, by an infected host. The well-known result is that $R_0$  has to exceed one for an epidemic outbreak to occur.  As pointed out in many studies, this prediction strictly holds only in deterministic  models with infinite population~\cite{hartfield2013introducing}. The underlying contact network also strongly influences the epidemic threshold and its predictions~\cite{pastor2001epidemic,wang2016predicting}. 

Furthermore, in finite populations, due to finite-size estimation challenges, an accurate identification of the epidemic threshold is problematic, and instead, an epidemic (critical) interval may be considered~\cite{erten2017criticality}. Following~\cite{lizier2012coherent,lizier2014framework}, the study of Erten et al.~\cite{erten2017criticality} applied an information-theoretic model of distributed computation to a homogeneous network. It identified (i) the lower bound of the interval, on the ordered side of the transition, by the peak of the active information storage, quantifying the ``memory'' of the computation during a contagion, and (ii) the upper bound of the interval, on the disordered side of criticality, with the maximum of the transfer entropy, quantifying the ``communication'' aspect of the contagion~\cite{erten2017criticality}.
 
Identifying and detecting critical regimes of a contagion remains a subject of a vigorous research.   The study of Hartfield and Alizon~\cite{hartfield2013introducing} contrasted the Critical Community Size (CCS), defined as the total population size needed to sustain an outbreak once it has appeared, with the Outbreak Threshold ($T_0$), computed at the onset of an outbreak, and measuring how many infected individuals are ``needed to ensure that an outbreak is very unlikely to go extinct by drift''. 

Under some circumstances (such as pathogen mutations or changes in the host population), even a maladapted pathogen with $R_0$ below but close to 1 still has a potential for an outbreak, when the changes cause its $R_0$ to exceed 1. This situation has been considered for new pathogens emerging by crossing the species barrier~\cite{antia2003role}. The sudden changes in the disease spread are obviously similar to critical dynamics and have been studied in this context, in an attempt to predict, and hopefully prevent,  the emergence of criticality~\cite{o2013theory}. Again, the underlying network topology and its mixing patterns can substantially affect the disease emergence~\cite{leventhal2015evolution}.

In summary, some of the present challenges relate to reliably detecting critical thresholds in finite-size systems, within complex network topologies, and dealing with distributed data generated by nonlinear dynamics.

A recent thermodynamics-based framework that has successfully dealt with such challenges in several abstract settings uses Fisher Information, a measure that  is directly connected to the rate of
change of the corresponding order parameters~\cite{wang2011fisher, prokopenko2011relating, crosato2018thermodynamics, kalloniatis2018}. These studies accurately identified phase transitions in the global spatiotemporal behaviour via an estimation of Fisher Information for a number of network topologies.  Critical thresholds have been pinpointed when the observed variables were most sensitive to the control parameters, resulting in divergence of Fisher Information in infinite systems and its maximisation in finite-size systems. Crucially, this method relies on estimation of underlying probability densities and, thus, is applicable even when the corresponding order parameter is unknown or cumbersome to compute.  

The approach based on Fisher Information has not, to date, been applied in an epidemiological setting, and promises to strengthen the prediction accuracy of epidemic thresholds in complex scenarios, involving heterogeneous network topologies, large-scale distributed data, and probable emerging pathogens. It may also enable derivations of closed form solutions in specific cases. To fully exploit its potential, however, the framework needs to be well grounded in a statistical-mechanical setting and complemented with rigorous methods for the suitable estimation of probability densities.

Typically, in statistical mechanics, the model of the systems is specified in full by providing the microscopic coupling constants between all components of the systems. Once the model is given, the challenge is to infer the emerging macroscopic properties of the ensemble using analytic and computational methods. An inverse problem, that is the determination of the microscopic coupling constants from some known macroscopic constraints is solved using the Maximum Entropy principle~(MaxEnt). The MaxEnt principle states that the least biased model is obtained by maximising the entropy of the distribution while at the same time respecting the imposed constraints. 

The MaxEnt principle has been applied to analyse activity in various complex networks including ecological networks, networks of neurons~\cite{tkavcik2013simplest}, biochemical and genetic networks and flocking birds~\cite{bialek2012statistical}. Typically, the solution of the variational Maximum Entropy problem relies on detailed computer simulations. Even in these typically hard cases, reasonable approximations can simplify the problem and lead to closed form solutions. Closed form solutions reveals strong analogies between macroscopic systems such as flock of birds and well understood statistically mechanical models such as the Ising model. These analogies elucidate the statistical mechanical origins of non-trivial collective phenomena, such as phase transitions, in complex systems.
  
Applications of the MaxEnt principle in computational epidemiology has so far been limited. The MaxEnt principle has been applied to stochastic SIS and SIR dynamic models, using real data to fit probability distributions of various epidemic characteristics such as time to infection, number of recovered individuals, number of infected individuals at a specific time interval [0,t], etc.~\cite{artalejo2011sis}. Going beyond this goal, we aim to apply the MaxEnt principle in construction of statistical mechanical models of epidemics enabling a statistical mechanical analysis of epidemic phase transitions. 

The state space of the individual nodes on an SIS epidemic network is binary since the nodes can be either infected or susceptible. This is reminiscent of the Ising model, where the state of each node is also binary (either up or down). However, unlike  the Ising model, the contact interactions in epidemic networks are directional. An infected individual has the capacity to flip the state of a susceptible neighbour while the susceptible individual has no effect on the infected neighbour. This directionality of interactions poses a novel challenge to the application of the MaxEnt method. 

In this paper, we apply the MaxEnt principle to derive a statistical mechanical model of a contact network undergoing SIS dynamics. We obtain an analytic solution for a simple case, characterising an initial (seeding) state of an outbreak. In order to arrive at analytic models for more general cases, we propose a simplification that assumes independence between the number of infected individuals and the number of infective links. We assess the impact of this assumption by comparing the results of the MaxEnt model of SIS process on Watts-Strogatz network~\cite{watts1998collective} to the results generated by computer simulations of the underlying dynamics. The derived MaxEnt models are used to evaluate quantities such as entropy, free energy, and Fisher Information in an epidemiological context. The statistical mechanical setting is used to provide a novel interpretation of epidemic thresholds. We specifically study the thermodynamic efficiency of contagion, considered as a distributed computational process. The thermodynamics of computation have recently been investigated in various contexts ~\cite{prokopenko2014transfer,spinney2016transfer,spinney2017transfer,crosato2018thermodynamics,kempes2017thermodynamic,spinney2018entropy}, but have not been applied to studies of epidemics.

The paper is structured as follows. In Background we describe relevant epidemic and network models, while Technical preliminaries outline the Maximum Entropy principle and an approach to criticality analysis based on Fisher Information. We then develop our framework applying the Maximum Entropy principle to an SIS epidemic model, including a closed form solution derived in specific cases. This is followed by computational results demonstrating criticality in statistical mechanical terms.

\section{Background}
\subsection{Models of epidemics}
The SIS model of epidemics captures the dynamics of diseases which are transmitted by individual to individual contact. Additionally, it refers to a type of disease in which individuals can be infected multiple times throughout their lives without developing long-lasting immunity. Examples of diseases following SIS dynamics are rotaviruses, sexually transmitted infections and bacterial infections~\cite{keeling2008modeling}. The SIS model of epidemics refers both to a differential equation model and a discrete time update process~\cite{pastor2001epidemiccn}.
In the case of the differential equation model of the SIS dynamics, the progression of the disease within the population is described by a pair of coupled ordinary differential equations.
\begin{equation}\label{SIS_S}
\frac{dS}{dt} = \gamma I - \beta IS 
\end{equation}
\begin{equation}\label{SIS_I}
\frac{dI}{dt} = \beta IS - \gamma I
\end{equation}

where, $I$ is the number of infected individuals, $S$ is the number of susceptible individuals, the parameter $\beta$ is the transmission rate and the parameter $\gamma$ is the recovery rate~\cite{keeling2008modeling}. This model assumes that all individuals within the host population interact with equal probability~\cite{keeling2008modeling} and is often referred to as the mass action model of infection. 

As the population is generally considered in isolation, it is also common to consider a single differential equation, normalising the population that is; \begin{equation}\label{SIS_I_only}
\frac{dI}{dt} = \beta I\left(1-I\right) - \gamma I
\end{equation}
\begin{equation}\label{norm_SIS}
S+I=1
\end{equation}
A well known result of Kermack and McKendrick~\cite{kermack1927contribution} shows that if the initial fraction of susceptibles is less than $\frac{\gamma}{\beta}$, $\frac{dI}{dt}<0$, the infection dies out. This is referred to as the "threshold phenomenon". This result can also be interpereted as requiring $\frac{\beta}{\gamma} = R_0$, commonly known as the basic reproductive ratio, to be large enough that the initial infected population increases with time. In a more general sense, the reproductive ratio $R_0$ is defined as ``A measure of the number of infections produced, on average, by an infected individual in the early stages of an epidemic when virtually all contacts are susceptible''~\cite{porta2014dictionary}. $R_0$ is frequently used in order to broadly quantify the transmissibility of an epidemic strain; in general, epidemics emerge when $R_0 > 1$~\cite{lloyd2005superspreading}.
 
\subsection{Network models}
Original approaches have assumed that interactions occur completely at random within the population~\cite{kermack1927contribution}. It has since been argued that  because the ``structure of a contact network can have a profound effect on the dynamics of infectious disease''~\cite{keeling2005implications} it is imperative to use network models as opposed to the more traditional mass-action models~\cite{kermack1927contribution}.
Thus networks have become a standard model for studying the spread of disease, quantifying interactions between individuals or populations of individuals~\cite{andersson1998limit, newman1999scaling, pastor2001epidemic, pastor2001epidemiccn,  keeling2005networks, newman2002spread}. 

There are a number of networks which are commonly investigated within the epidemiological literature. The most commonly studied have been random networks~\cite{andersson1998limit}, lattices~\cite{rhodes1996dynamics}, scale-free networks~\cite{pastor2001epidemic} and small world networks~\cite{pastor2001epidemiccn}. 

Unlike the mass action mixing approach, network based approaches define a neighbourhood for each individual in which they can infect others and be infected by others. Importantly, in some cases this representation yields a closed form solution, for example the study of~\citep{newman2002spread} shows that a large class of the SIR models of epidemic disease can be solved exactly on networks of various kinds using a combination of percolation models and generating function methods. Another key distinguishing result in the study of epidemics on networks shows that in some cases there is no critical threshold~\cite{pastor2001epidemic}, with disease propagating regardless of the probability of infection~\cite{pastor2001epidemic}. As such, for network models there is no general result analogous to $R_0 = 1$ from differential equations models.

Of special importance to studies of contagion processes is the class of small-world networks introduced by Watts and Strogatz~\cite{watts1998collective}. The algorithm constructing a small world network essentially interpolates between regular and random networks, beginning with a  lattice, and rewiring edges with a given probability. The small world networks are characterised by a high clustering coefficient and
small average path length. The clustering coefficient considers each vertex of the graph individually, and compares the number of edges between neighbours to a complete graph as a ratio, while the average path length is the average minimum distance between two vertices. The Watts-Strogatz model~\cite{watts1998collective}, produces a graph with a small average path length, and a resulting clustering coefficient which is significantly higher than the corresponding coefficient of an Erdos-Renyi random graph model~\cite{erdos1959random}. Importantly, networks of different topologies may be considered as small-world networks as long as they are characterised by a relatively high clustering coefficient and a relatively low average path length. These features are of particular relevance to studies of contagion in many real world scenarios. The degree distribution for a Watts-Strogatz network,  interpolating between the ring lattice and a random graph, is similar to the distribution of a random graph but has a pronounced peak centred on the mean degree, decaying exponentially for degrees deviating from the mean ~\cite{barrat2000properties}.

Typically in SIS discrete time update models, the infection parameter $\nu$ defines a per contact (edge) per time--step probability of transmission. That is, given an individual $x_i$ in a neighbourhood with $r$ infected individuals, the per time step probability of infection $P(x_i)$~\cite{pastor2001epidemiccn} is 
\begin{equation}\label{prob_infection}
P(x_i) = 1-(1-\nu)^{r} .
\end{equation}

The parameter $\nu$ is analogous to the parameter $\beta$ in the differential equations model, however, there is a subtle difference in interpretation: $\beta$ is a continuous rate of transmission while $\nu$ is a discrete probability of transmission per time step.
As the update scheme is parallel, individuals all change states at the same time i.e. recovery events from the current time step cannot affect infection on the same time step and vice versa. 
Given our objective of studying criticality in a complex distributed setting via the Maximum Entropy principle, the various network topologies provide a natural constraint on the testable information with respect to interactions within the population. This constraint imposed by heterogeneous networks presents the key challenge in obtaining a closed form solution, unlike approaches based on mean field approximations.

\section{Technical Preliminaries}
\subsection{\label{MEM}The Maximum entropy method}
Often we are faced with the problem of determining the least biased probability distribution, consistent with a set of specific constraints on the average values of measurable quantities. These may, for example, represent relevant conserved quantities in a thermodynamic system, or generic constraints in any probabilistic system of multinomial form (i.e. a system composed of a number of distinguishable entities allocated to equiprobable distinguishable categories)~\cite{niven2010minimization}. In the most general setting this problem can be resolved by extracting the highest amount of (Shannon) information available. As pointed out by Jaynes, ``in making inferences on the basis of partial information we must use that probability distribution which has maximal entropy subject to whatever is known. This is the only unbiased assignment we can make''~\cite{jaynes1957information}.

The Shannon entropy $S$ of a discrete random variable  $A$ with state space $\mathcal{A} =\{\vec{\sigma}_1,\vec{\sigma}_2,\vec{\sigma}_3,...\}$, the set of all possible states, is given by~\cite{shannon2001mathematical}
\begin{equation}\label{Shan_ent}
 S(A)  = \sum_{\vec{\sigma} \in \mathcal{A}} -P(\vec{\sigma})\log(P(\vec{\sigma})) ,
\end{equation} 
where $P(\vec{\sigma})$ is the probability that the system is in the state $\vec{\sigma}$.

In order to extract the least biased probability distribution, one typically maximises the Shannon entropy~(\ref{Shan_ent}), subject to the normalisation and $K$ moment constraints on the system, shaped by some functions $f_k(\vec{\sigma})$ with measurable expectations $\langle f_k \rangle$, for $k = 1,\ldots,K$: 
\begin{equation}\label{Normalisation}
 \sum_{\vec{\sigma} \in \mathcal{A}} P(\vec{\sigma}) = 1
\end{equation}
and
\begin{equation}\label{conserved_quantity}
\sum_{\vec{\sigma} \in \mathcal{A}} P(\vec{\sigma}) f_k(\vec{\sigma}) = \langle f_k \rangle.
\end{equation}

In practice, the problem is an optimisation problem which can be solved using the method of Lagrange multipliers. Using the method of Lagrange multipliers, the form of the distribution which maximises the entropy is
\begin{equation}\label{most_prob}
P(\vec{\sigma}) =  \exp\left(-\lambda_0 -\sum_{k=1}^K \lambda_k f_k(\vec{\sigma})\right)
\end{equation}
where $\lambda \equiv \{\lambda_1, \ldots, \lambda_k, \ldots, \lambda_K \}$ is the set of Lagrange multipliers corresponding to $K$ constraints and $\lambda_0$ is the "Massieu function", the Lagrange multiplier corresponding to the normalisation constraint. Introducing the generalised partition function $Z(\lambda)=\exp{\lambda_0}$ yields
\begin{equation}\label{most_prob_Z}
P(\vec{\sigma}) =  Z(\lambda)^{-1}\exp\left({-\sum_{k=1}^K \lambda_k f_k(\vec{\sigma})}\right)
\end{equation}

\subsection{\label{fisher_tech_prelim}Fisher Information}

The Fisher Information is a measure of the information that an observable random variable $X$ contains about a set of unknown parameters $\lambda$, defined as
\begin{equation}\label{fisher}
F_X(\lambda) = E\left[ \left( \frac{\partial}{\partial \lambda} \log P(x;\lambda)   \right)^2 \right],
\end{equation}
which for continuous random variables is
\begin{equation}\label{fisherequivalent}
F_X(\lambda) = \int \left( \frac{\partial}{\partial \lambda} \log P(x;\lambda)   \right)^2 P(x;\lambda) dx,
\end{equation}
where $P(x;\lambda)$ is the probability density function~(pdf) of $X$ conditional on the parameters $\lambda$.

For a joint random variable, the Fisher Information has a chain rule decomposition~\cite{zamir1998proof} such that if $X$ and $Y$ are jointly distributed random variables,
\begin{equation}\label{fisher_chain}
F_{X,Y}(\lambda) = F_X(\lambda) + F_{Y|X}(\lambda)
\end{equation}

If $X$ and $Y$ are independent random variables, the distribution of $Y$ given $X$ is the same as the distribution of $Y$, and therefore, $F_{Y|X} = F_{Y}$ implying that
\begin{equation}\label{fisher_chain_independence}
F_{X,Y}(\lambda) = F_X(\lambda) + F_Y(\lambda).
\end{equation}

Often, it is important to reparametrise the Fisher Information~\cite{lehmann2003theory}:
\begin{equation}\label{Fisher_reparam}
F_X(\mu) = \left( \frac{d\lambda}{d \mu} \right)^2 F_X(\lambda(\mu)),
\end{equation}
where $\lambda$ and $\mu$ are both parametrisations of $X$, and $\lambda$ is a continuously differentiable function of $\mu$.

For many distributions the Fisher Information is known exactly, and in particular, we shall use the closed form representation for the Fisher Information of a Binomial$(n,q)$ random variable $\chi$:
\begin{equation}\label{fisher_binomial}
F_\chi(q) = \frac{n}{q(1-q)}.
\end{equation}

\subsection{\label{therm_eff_comp}Thermodynamic efficiency of computation}

In a statistical mechanical setting, for thermodynamic variables $\lambda$, the solutions obtained according to the Maximum Entropy principle are characterised by probability densities in the form of the Gibbs measure:
\begin{equation}
\label{eq:gibbs-measure}
P(\sigma|\lambda) = \frac{1}{Z(\lambda)}e^{-\beta H(\sigma,\lambda)} = \frac{1}{Z(\lambda)}e^{-\sum_k \lambda_k f_k(\sigma)} ,
\end{equation}
where the state functions $f_k(\sigma)$ are defined over the configuration space, $\beta=1/k_bT$ is the inverse temperature $T$ ($k_B$ is the Boltzmann constant), and the Hamiltonian $H(\sigma,\lambda)$ defines the total energy at state $\sigma$~\cite{brody1995geometrical, crooks2007measuring}. In other words, equation~\eqref{eq:gibbs-measure} expresses the state probability in terms of the state energy.

The Gibbs free energy of such system is given by:
\begin{equation}
\label{eq:gibbs-potential}
G(T,\lambda_m) = U(S,\phi_m) - TS -  \phi_m \lambda_m ,
\end{equation}
where $U$ is the internal energy of the system, $S$ is the configuration entropy and $\phi_m$ is an order parameter.
In such a setting, the Fisher Information quantifies the size of the fluctuations around equilibrium in the collective variables $f_m$ and $f_n$, and is proportional to the curvature of the free entropy $\psi = \ln Z = -\beta G$~\cite{brody1995geometrical, brody2003information, janke2004information, crooks2007measuring}:
\begin{equation}
 F_{mn}(\lambda) = \Big\langle (X_m(x) - \langle X_m \rangle ) (X_n(x) - \langle X_n \rangle ) \Big\rangle  = \frac{\partial^2\psi}{\partial\lambda_m\partial\lambda_n} .
\end{equation}
It also identifies phase transitions and the corresponding critical thresholds~\cite{wang2011fisher}, being proportional to the derivatives of the corresponding order parameters with respect to the thermodynamic variables $\lambda$~\cite{prokopenko2011relating}:
\begin{equation}\label{fisher_relation}
 F_{mn}(\lambda)  = \beta\frac{\partial\phi_m}{\partial\lambda_n} .
\end{equation} 
Furthermore, under a quasi-static protocol, the Fisher Information can be interpreted as the generalised work  $W_{gen}$~\cite{crosato2018thermodynamics}:
\begin{equation}
\label{eq:fisher-work-curvature}
F(\lambda) = - \frac{d^2\langle\beta W_{gen}\rangle}{d\lambda^2} .
\end{equation}

Using  equation~\eqref{eq:fisher-work-curvature}, the rate of expended work can be expressed as follows:
\begin{equation}\label{work_deriv}
{d \langle \beta W_{gen}\rangle}{/d\nu} = -\int_{\lambda^*}^{\lambda} F(\lambda') d\lambda' ,
\end{equation}
where $\lambda^*$ is the zero-response point for which small changes in the control parameter incur no work:
\begin{equation}\label{0rp}
\left. \frac{d\langle \beta W_{gen}\rangle}{d\lambda}\right|_{\lambda = \lambda^{*}} = 0 .
\end{equation}

Having determined, via Fisher information, the rate of expended work carried out to generate order within the system, one may define the thermodynamic efficiency of computation~\cite{crosato2018thermodynamics}, as ``the reduction in uncertainty (i.e., the increase in order) from an expenditure of work given a value of the control parameter'':
\begin{equation}\label{TEC}
\eta = \frac{-dS/d \lambda}{d \langle \beta W_{gen}\rangle/d\lambda}.
\end{equation}
In this work we shall extend the notion of the thermodynamic efficiency of computation to contagion processes.

\section{Maximum entropy framework for epidemics}
\subsection{A network model of SIS epidemic}
We will consider a graph $G(\cal{V},\cal{E})$ with vertex set $\cal{V}$ and edge set $\cal{E}$. The nodes $i \in {\cal{V}}  = \{ 1,2,3 \ldots \}$ represent individuals in the population taking one of two states: susceptible or infected. The state of vertex $i$ will be denoted $\sigma_i$ and will take value 0 if the individual is susceptible or 1 if the individual is infected. The edges ${(i,j)} \in \cal{E}$ represent connections between two individuals $i$ and $j$ along which infection can spread.
The state of the entire system, $\vec{\sigma}$ will be expressed as the vector comprising the states of all individuals. That is, the $i$th element of $\vec{\sigma}$ is $\sigma_i$. We will denote the set of all states $\vec{\sigma}$ as $\mathcal{A}$.

The epidemic dynamics which we will investigate within the maximum entropy framework are SIS epidemics spreading on a network. The analytical results will be contrasted with the simulation results obtained by stochastic discrete-time parallel-update SIS model introduced by~\cite{pastor2001epidemiccn}. Similar to the deterministic SIS model, recovery of individuals in this model occurs independently of their neighbours, with a constant probability of recovery, denoted $\delta$. This is analogous to the parameter $\gamma$ from the SIS differential equation model, with the difference being that $\gamma$ represents a rate, whereas $\delta$ is a probability per discrete time step. 

We aim to investigate the maximum entropy distribution corresponding to a SIS epidemic spreading on a graph $G(\cal{V},\cal{E})$, constrained by the testable information formed by the averages of two variables in the steady state of the SIS dynamics. These  averages, thermodynamically corresponding to conserved quantities, are defined as follows:
\begin{equation} \label{I}
I(\vec{\sigma}) = \sum_{\sigma_i \in \vec{\sigma}}\sigma_i  
\end{equation}
\begin{equation} \label{C}
C(\vec{\sigma}) = \sum_{\sigma_i \in \vec{\sigma}} \sigma_i \sum_{j \in N_i} 1-\sigma_j.
\end{equation}
where $N_i$ denotes the neighbourhood of node $i$. The quantity $I(\vec{\sigma})$ is the total number of infected individuals in a configuration $\vec{\sigma}$ and is of clear interest in the study of infectious diseases. The quantity $C(\vec{\sigma})$ is the number of neighbouring individuals who have opposite states. In the context of epidemic modeling, this is the number of potentially infective connections in a configuration $\vec{\sigma}$. In analogy to an electromagnetic spin model such as the Ising model, the potentially infective connections ($C$) correspond to the node-node interaction energy, whereas the number of infected individuals ($I$) corresponds to the energy due to the applied external magnetic field.

The SIS discrete-time update dynamics have both an equilibrium state (i.e., the absorbing state of the system) and a metastable state (i.e., a state which takes time exponential in the number of vertices to leave). The equilibrium state corresponds to the trivial case in which $I=0$. At the metastable state, however, the rate at which infected individuals are recovering and the rate at which susceptible individuals are being infected are equal. Between each time step, the rate of recovery is proportional to $I$, whereas the instantaneous rate of infection is proportional to $C$. 

\subsection{Maximising Entropy}
In order to obtain the most likely distribution of infection within the simulated population during a steady state, we use the maximum entropy method with constraints on the average value of $I$ as given by equation~(\ref{I}) and $C$ as given by equation~(\ref{C}). Formally,  the Maximum Entropy principle for this system with SIS discrete time update dynamics forms the optimisation problem
\begin{equation}\label{Formal_def_of_problem}
\max_{P(\vec{\sigma})} S(A) \textrm{, for } S(A) =  -\sum_{\vec{\sigma}{\in \mathcal{A}}} {P}(\vec{\sigma})\log({P}(\vec{\sigma})) ,
\end{equation}
subject to
\begin{equation}\label{expectedI}
\langle I \rangle = \sum_{\vec{\sigma} \in \mathcal{A}}\bigg{(}P(\vec{\sigma})\sum_{\sigma_i \in \vec{\sigma}} \sigma_i\bigg{)}
\end{equation}
\begin{equation}\label{expectedC}
\langle C \rangle = \sum_{\vec{\sigma} \in \mathcal{A}}\bigg{(}P(\vec{\sigma})\sum_{\sigma_i \in \vec{\sigma}} \sigma_i \bigg{(}\sum_{j \in N_i} 1-\sigma_j\bigg{)}\bigg{)}
\end{equation}
\begin{equation}\label{Normalisation_SIS}
1 = \sum_{\vec{\sigma} \in \mathcal{A}} P(\vec{\sigma})
\end{equation}
The form of this MaxEnt solution for $K$ constraints is given by equation~(\ref{most_prob_Z}) specifying Gibbs distribution. The specific Gibbs distribution consistent with the constraints given by equations~(\ref{expectedI}), (\ref{expectedC}) and (\ref{Normalisation_SIS}) is
\begin{equation}\label{Prob_ME}
P(\vec{\sigma}) = \frac{e^{\lambda_1 I(\vec{\sigma})+\lambda_2 C(\vec{\sigma})}}{Z}
\end{equation}
where $I(\vec{\sigma})$ and $C(\vec{\sigma})$ are  defined by (\ref{I}) and (\ref{C}) respectively, and $\lambda_1, \lambda_2$ and $Z$ are unknown Lagrange multipliers. Thus, in order to solve for the Lagrange multipliers and obtain the maximum entropy distribution consistent with known information, one must use the  the  averages of $I$ and $C$.

For clarity, we abbreviate
\begin{equation}\label{substitutions}
x = e^{\lambda_1}, \hspace*{10pt} y = e^{\lambda_2}.
\end{equation}
Substituting~\eqref{substitutions} into \eqref{expectedI}--\eqref{Prob_ME}, we obtain
\begin{equation}\label{Prob_ME2}
P(\vec{\sigma}) = \frac{x^{I(\vec{\sigma})}y^{C(\vec{\sigma})}}{Z}
\end{equation}
\begin{equation}\label{expectedI2}
\langle I \rangle = \sum_{\vec{\sigma} \in \mathcal{A}} I(\vec{\sigma})\frac{x^{I(\vec{\sigma})}y^{C(\vec{\sigma})}}{Z}
\end{equation}
\begin{equation}\label{expectedC2}
\langle C \rangle = \sum_{\vec{\sigma} \in \mathcal{A}} C(\vec{\sigma})\frac{x^{I(\vec{\sigma})}y^{C(\vec{\sigma})}}{Z}
\end{equation}
\begin{equation}\label{Normalisation2}
1 = \sum_{\vec{\sigma} \in \mathcal{A}} \frac{x^{I(\vec{\sigma})}y^{C(\vec{\sigma})}}{Z}
\end{equation}
Let us define the sets $\aleph_{I,C}$, the sets of all configurations $\vec{\sigma}$ such that the total number of infected individuals $I(\vec{\sigma})=I$, and the number of potentially infective connections $C(\vec{\sigma})=C$; formally, we have
\begin{equation}\label{aleph}
\aleph_{I,C} =\{\vec{\sigma}: I(\vec{\sigma}) = I, C(\vec{\sigma}) = C\}.
\end{equation}
 The sets $\aleph_{I,C}$ form a partition of $\mathcal{A}$, the space of all configurations. Hence, each configuration $\vec{\sigma}$ belongs to exactly one of the sets $\aleph_{I,C}$.  As all elements of the set $\aleph_{I,C}$ have the same value of $I(\vec{\sigma})$ and $C(\vec{\sigma})$, we see from equation~(\ref{Prob_ME2}) that all states which belong to the same set $\aleph_{I,C}$  have identical probability.  Specifically, the probability of a state $\vec{\sigma} \in \aleph_{I,C}$ is
\begin{equation}\label{P_IC}  
P(\vec{\sigma} \in \aleph_{I,C}) = \frac{x^{I}y^{C}}{Z}
\end{equation}
As each of the states $\vec{\sigma}$ in a set $\aleph_{i,c}$ has identical probability, this naturally follows the concept of a macrostate $(I,C)$. By denoting the cardinality of the set $\aleph_{I,C}$ by $N(I,C)$, we  simplify the  system of equations~(\ref{expectedI2})--(\ref{Normalisation2}) as
\begin{equation}\label{expectedI3}
\langle I \rangle = \sum_{I,C} N(I,C) I \frac{x^I y^C}{Z}
\end{equation}
\begin{equation}\label{expectedC3}
\langle C \rangle = \sum_{I,C} N(I,C) C \frac{x^I y^C}{Z}
\end{equation}
\begin{equation}\label{Normalisation3}
1 = \sum_{I,C} N(I,C) \frac{x^I y^C}{Z}
\end{equation}
This reformulation opens a way to MaxEnt solutions expressed in terms of the probabilities $P(I,C)$ defined over the macrostates $(I,C)$.  


\subsection{Example: Numerical solutions for small graphs}\label{complete_graph_ex}

The MaxEnt equations (\ref{expectedI3})-(\ref{Normalisation3}) is a system of polynomial equations in $x$, $y$ and $Z$, which in general do not have an analytic solution. The solutions to these equations can be obtained numerically as we will demonstrate in this section. 

\begin{figure}
\begin{subfigure}{0.2\textwidth}
   \includegraphics[width=1\linewidth]{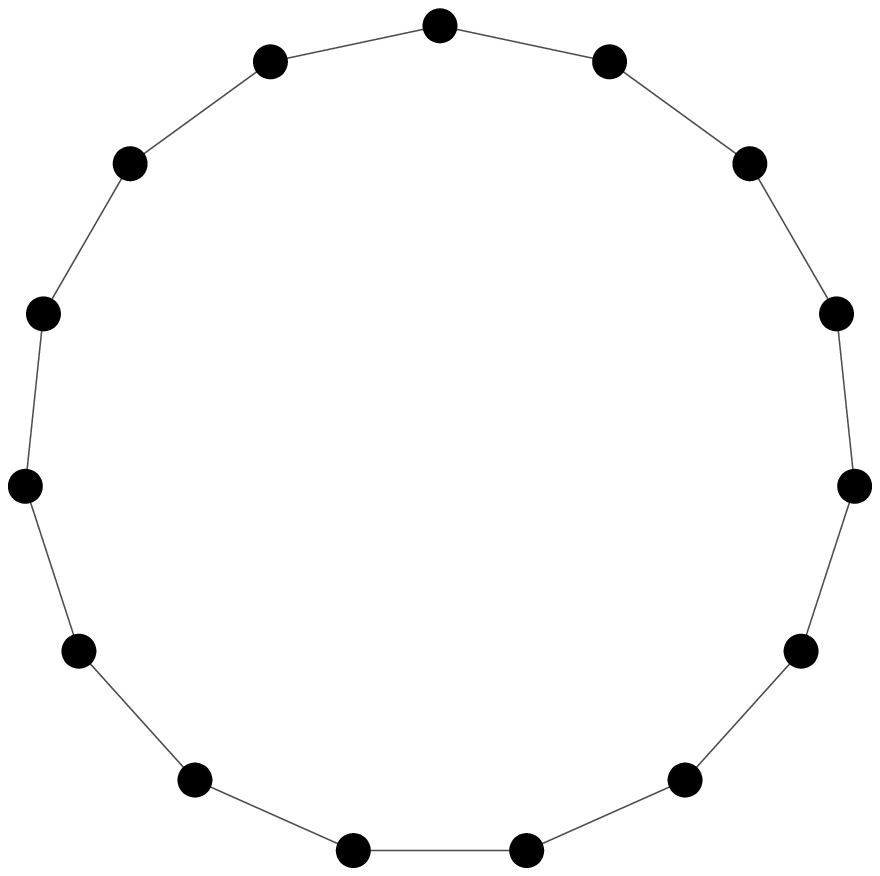}
   \caption{}
   \label{fig:ring_network} 
\end{subfigure}
\begin{subfigure}{0.2\textwidth}
   \includegraphics[width=1\linewidth]{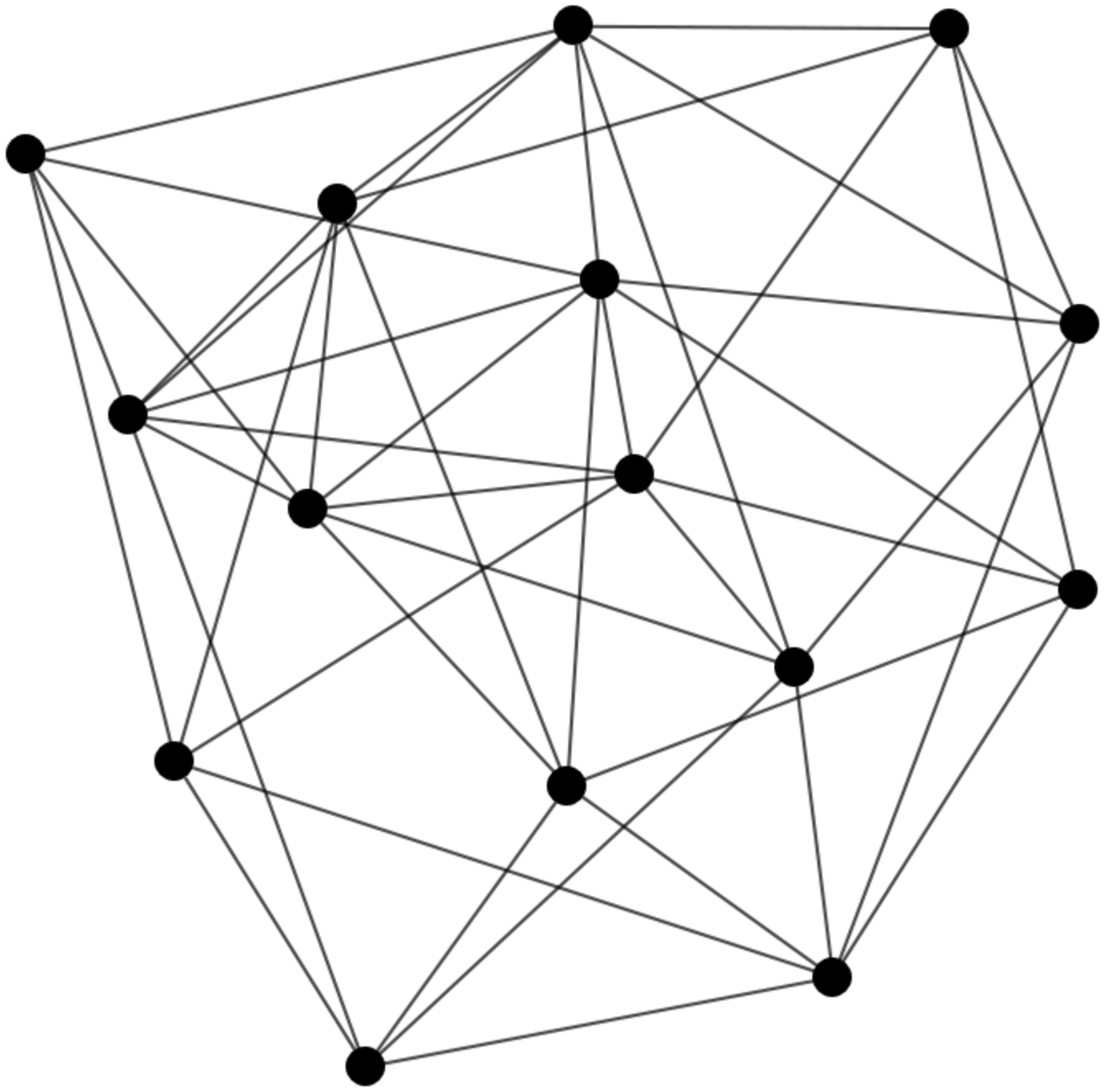}
   \caption{}
   \label{fig:random_network}
\end{subfigure}

\caption[Two example networks]{Example networks of 15 nodes for numerically testing the MaxEnt model. a) a ring network b) Wattz-Strogatz network with $p=1$ and $k=6$.}
\label{fig:small_networks}
\end{figure}

To exemplify the numerical construction of our MaxEnt model of SIS epidemics on complex networks, we have chosen two graph topologies shown in Figure \ref{fig:small_networks}. The first graph is a ring and the second graph is a Watts-Strogatz random graph with $p=1$ and $k=6$. Both graphs consists of 15 nodes. Since these networks are small it is straightforward to compute the functions $N(I,C)$ by listing all possible state configurations and counting the number of configurations with specific values of $I$ and $C$. Such method of computing $N(I,C)$ is feasible for small networks, but for larger networks one should should make use of more sophisticated combinatorial algorithms.

In practical applications, the values of constraints, $\langle I \rangle$ and $\langle C \rangle$, would be taken from field data. For our purposes, we can obtain the ``data" by computationally simulating the dynamics of an SIS epidemic. We simulate the dynamics using SIS parallel update processes stochastic simulation with an arbitrary chosen values of probability of infection transmission $\nu$ and probability of recovery $\delta$. The simulation involves initializing the system at random and then updating the state of the system for 120000 timesteps. The first 20000 timesteps are used to equilibrate the system and the final 100000 timesteps are used for sampling the equilibrium state. At every timestep, the number of infected individuals $I$ and number of infected connections $C$ is recorded, and this is used to compute $\langle I \rangle$ and $\langle C \rangle$. The values of $\langle I \rangle$ and $\langle C \rangle$ are then used as constraints in equations (\ref{expectedI3})-(\ref{expectedC3}). The equations (\ref{expectedI3})-(\ref{Normalisation3}) are solved numerically for $x$, $y$ and $Z$.

\begin{figure*}
\includegraphics[scale=1]{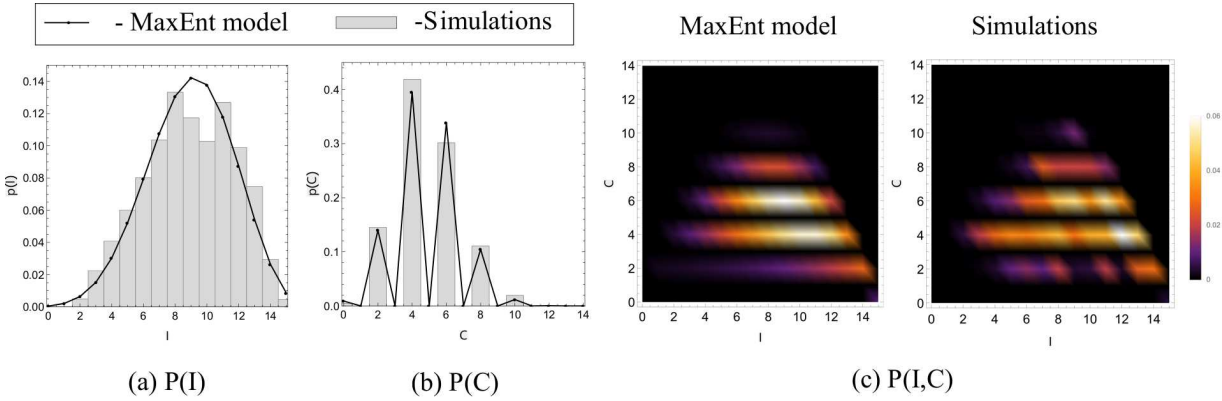}

\caption{The observed simulation and MaxEnt probability density functions $P(I)$, $P(C)$ and $P(I,C)$ for a 15 node ring graph (Figure \ref{fig:small_networks}(a)). The SIS simulation parameters were $\nu = 1.0\times10^{-3}$, $\delta = 6.0\times 10^{-4}$. The empirical values of the constraints for the MaxEnt equations were $\langle I \rangle = 8.9323$ and $\langle C \rangle = 4.8542$. The computed Lagrange multipliers were $\lambda_1 = 0.1938$ and $\lambda_2 = -0.6957$.}
\label{fig:ring_network_p}
\end{figure*}

\begin{figure*}
\includegraphics[scale=1]{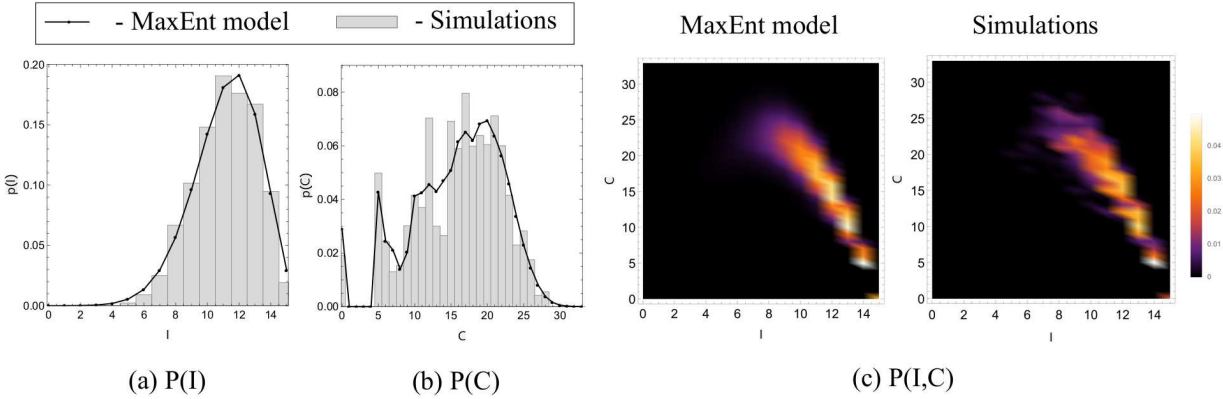}
\caption{The observed simulation and MaxEnt probability density functions $P(I)$, $P(C)$ and $P(I,C)$ for a 15 node random graph (Figure \ref{fig:small_networks}(b)). The SIS simulation parameters were $\nu = 3.288\times10^{-4}$ and $\delta = 5.0\times 10^{-4}$. The empirical values of the constraints for the MaxEnt equations were $\langle I \rangle = 11.1624$ and $\langle C \rangle = 15.8917$. The computed Lagrange multipliers were $\lambda_1 = 0.6541$ and $\lambda_2 = -0.1500$.}
\label{fig:random_network_p}
\end{figure*}

Figure~\ref{fig:ring_network_p} and Figure~\ref{fig:random_network_p} compare the computed MaxEnt probability distributions with distributions from simulation data for the ring and random networks. There is a good agreement between the empirical and MaxEnt distributions $P(I)$, $P(C)$ and $P(I,C)$ for both ring and random network topologies supporting the validity of the MaxEnt method. 

The numerical solutions of MaxEnt equations is of general applicability. However, in order to arrive to epidemiological interpretation of the Lagrange multipliers $\lambda_1$ and $\lambda_2$, we would like to express them analytically in terms of $\langle I \rangle$ and $\langle C \rangle$. The analytic solution may only be obtained after invoking certain simplifying assumptions, in particular, independence between $P(I)$ and $P(C)$. The derivation and the analysis of the analytic solutions is presented in the following sections.

\subsection{Example: Complete graph}\label{complete_graph}

We will start our analytic analysis by considering 
the specific case in which $G(\cal{V},\cal{E})$ is a complete graph, with $V$ vertices and $E = \frac{V(V-1)}{2}$ edges. Given $I$ infected individuals, for each infected node there are $V-I$ susceptible neighbours.  Therefore, there are $C = I(V-I)$ potentially infective connections. It is worth noting that although a given number of infected individuals on this topology defines precisely the number of potentially infective connections, the converse is not true in general. That is, for a number of potentially infective connections there generally more than one corresponding number of infected individuals.

Each $I$ specifies the value of $C$, however each value of $C$ often defines exactly two values of $I$. As knowing $I$ uniquely defines $C$, $\aleph_{I,C} = \aleph_{I} =\{\vec{\sigma}: I(\vec{\sigma}) = I\}$. Consequently, $N(I,C) = N(I)$. In a complete graph with $V$ vertices, $N(I) = {V \choose I}$,  analogous to the outcome of placing $I$ balls into $V$ buckets, resulting in the binomial coefficient. Therefore,  the total number of infected individuals, represented by the constraint (\ref{expectedI3}), is a sum from $I = 0$ to $V$, giving
\begin{equation}\label{expectedI4}
\langle I \rangle = \sum_{I=0}^{V} {V \choose I} I \frac{x^I y^{I(V-I)}}{Z}
\end{equation}
Similarly, the constraint (\ref{expectedC3}) yields:
\begin{equation}\label{expectedC4}
\bigg{\langle} I(V-I) \bigg{\rangle} = \sum_{I=0}^{V} {V \choose I} I(V-I) \frac{x^I y^{I(V-I)}}{Z}
\end{equation}
while the normalisation constraint (\ref{Normalisation3}) becomes:
\begin{equation}\label{Normalisation4}
1 = \sum_{I=0}^{V} {V \choose I}\frac{x^I y^{I(V-I)}}{Z}
\end{equation}
This is not analytically reducible further, and in the next subsection we consider a simplified system where an exact solution can be found.

Henceforth, for a graph with $V$ vertices and $E$ edges, we shall use average quantities $\langle I^{*} \rangle = \frac{\langle I \rangle}{V}$ and $\langle C^{*} \rangle = \frac{\langle C \rangle}{E}$.
 
\subsection{Example: An initial seeding state}\label{simplified}
In order to illustrate how the Maximum Entropy principle yields an analytical solution, we consider a very simple system with two constraints: on the number of infected individuals $I$, and the usual normalisation constraint. This abridged case (not limited to the complete graph topology) describes the precursor state of the epidemic with a number of infection sources distributed within the network.  This state essentially corresponds to the initial state of an outbreak, before any infective transmissions have taken place, and agent-based simulation studies often focus on such initial ``seeding'' state in quantifying effects of different seeding scenarios \cite{germann2006mitigation,cliff2018,zachreson2018}. Nevertheless, this case will reveal an important thermodynamic analogy between  key variables of SIS model and the inverse temperature of Gibbs distribution resulting from entropy maximisation.
Explicitly stated, we wish to find the Maximum Entropy solution in the form \begin{equation}\label{Prob_ME_1var}
P(\vec{\sigma}) = \frac{e^{\lambda I(\vec{\sigma})}}{Z}
\end{equation}
subject to the following constraints: 
\begin{equation}\label{avI}
 \frac{1}{Z} \sum_{I = 0}^V I \binom{V}{I} x^I =\langle I \rangle
\end{equation}
\begin{equation}\label{norm}
\frac{1}{Z} \sum_{I = 0}^V \binom{V}{I} x^I = 1
\end{equation}
The binomial theorem transforms \eqref{norm} into
\begin{equation}\label{normalisation_1_var}
Z = (x+1)^V
\end{equation}
where $x = e^\lambda$ is a positive real number.
Substituting (\ref{normalisation_1_var}) into (\ref{avI}) yields:
\begin{equation}\label{directway}
\langle I \rangle = \frac{\sum_{I = 0}^V I \binom{V}{I} x^I}{(x+1)^V}  = \frac{Vx(x+1)^{V-1}}{(x+1)^V} = \frac{Vx}{(x+1)}
\end{equation}
One may be interested in tracing the proportion of the infected individuals within the total population, $\langle I^{*} \rangle = \frac{\langle I \rangle}{V}$, which is immediately obtained from (\ref{directway}): 
\begin{equation}\label{x_solution_1_variable}
\langle I^{*} \rangle = \frac{x}{x+1}
\end{equation}
Hence, the Lagrange multiplier corresponding to the constraint (\ref{avI}), $\lambda = \log x$,  is
\[ \lambda = \log \frac{\langle I^{*} \rangle}{1-\langle I^{*} \rangle}  \]
Finally, substituting this expression into the solution (\ref{Prob_ME_1var}), we obtain the Gibbs distribution:
\begin{equation}\label{psigma}
P(\sigma) = \frac{e^{ \bigg{(}\log \frac{\langle I^{*} \rangle}{1-\langle I^{*} \rangle}\bigg{)}  I(\sigma)}}{Z}.
\end{equation}
Thus, interpreting  $I(\sigma)$  as the energy of the system, we may derive a thermodynamic analogy of the equilibrium inverse temperature in the SIS epidemic model as negative $\lambda$, that is, $\beta = \log \frac{ 1-\langle I^{*} \rangle }{\langle I^{*} \rangle}$, i.e., the log-ratio between non-infected and infected proportions during the initial state of an outbreak.

The partition function, i.e., the Lagrange multiplier corresponding to the normalisation constraint (\ref{norm}), can be explicitly resolved by using (\ref{x_solution_1_variable}) and~(\ref{normalisation_1_var}):
\[Z = \bigg{(}\frac{1}{1-\langle I^{*} \rangle}\bigg{)}^V\]
Therefore, 
\begin{equation}\label{exacto2}
P(I) = \binom{V}{I} \langle I^{*} \rangle^I  \left (1-\langle I^{*} \rangle \right)^{V-I}
\end{equation}
This is the binomial distribution with the ``probability of success''  $\langle I^{*} \rangle$ during $V$ trials, characterising the initial state of an epidemic outbreak.

\subsection{Assuming independence}
In order to obtain an analytical solution to the more general problem that includes, in addition, the constraint on the number of infective links (\ref{expectedC}), we introduce an assumption about the the random variables $I$ and $C$. Specifically, we assume that 
$I$ and $C$ are independent: $P(I,C) = P(I)P(C)$, and as a result, $N(I,C) = N(I)N(C)$.

Under this assumption, the average of $I$ may only be used to infer the distribution of $I$. Similarly, the average of $C$ may only be used to infer the distribution of $C$. This allows us to reduce the constraints (\ref{expectedI3})--(\ref{Normalisation3}) as follows:
\begin{equation}\label{I_indep}
\langle I \rangle = \sum_{I \in V} I N(I) \frac{x^{I} }{Z_I}
\end{equation}
\begin{equation}\label{C_indep}
\langle C \rangle = \sum_{C \in E} C N(C) \frac{y^{C}}{Z_C}
\end{equation}
\begin{equation}\label{I_Normalisation}
1 = \sum_{I \in V}  \frac{x^{I} }{Z_I}
\end{equation}
\begin{equation}\label{C_Normalisation}
1 = \sum_{ C \in E} \frac{y^{C}}{Z_C}
\end{equation}
where $Z_C  Z_I = Z$.
We then follow the derivations outlined in subsections \ref{complete_graph} and \ref{simplified}, solving (\ref{I_indep}) and (\ref{I_Normalisation}) for $P(I)=\frac{x^{I} }{Z_I}$, and (\ref{C_indep}) and (\ref{C_Normalisation}) for $P(C)=\frac{y^{C}}{Z_C}$ separately. Doing so, we find that $P(I)$ is a binomial random variable with $V$ trials and the probability of success $\langle I^* \rangle$ and that $P(C)$ is a binomial random variable with $E$ trials and the probability of success $\langle C^* \rangle$.
Thus, our next result is that
\begin{multline}\label{binom_pic}
 P(I,C) = \\
 \binom{V}{I}\binom{E}{C} \langle I^* \rangle^I(1-\langle I^* \rangle)^{V-I} \langle C^* \rangle^C(1-\langle C^* \rangle)^{E-C} 
\end{multline}
Under the assumption that $I$ and $C$ are independent, we may express the entropy of the system in its steady state, that is, the entropy of the joint variable $(I,C)$, as the sum of the individual entropies of $I$ and $C$: 
\[ H(I,C) = H(I)+H(C) \]

Furthermore, given the partition functions of the marginal probability distributions obtained as 
\[ Z_I = \bigg{(}\frac{1}{1-\langle I \rangle^{*}}\bigg{)}^V  \textrm{ and } Z_C = \bigg{(}\frac{1}{1-\langle C \rangle^{*}}\bigg{)}^E  \]
we derive the free entropy of the entire system as
\begin{multline}
\label{free_entropy2}
\log(Z) = \log(Z_I  Z_C) = \log(Z_I)+\log(Z_C) = \\
-V \log(1-\langle I \rangle^{*}) -E \log(1-\langle C \rangle^{*})  
\end{multline}

\subsection{Fisher Information}

Now we will use the derived Gibbs distributions in expressing the Fisher Information of the joint random variable $(I,C)$ and characterising critical regimes of SIS epidemics. Under the assumption that $I$ and $C$ are independent, $F_{I,C}(\nu) = F_I(\nu) + F_C(\nu)$~\cite{zamir1998proof}. 

Our control parameter in this case is $\nu$, and the Fisher Information of the joint variable $(I,C)$ describing the macroscopic state of the system will be derived with respect to $\nu$, and then with respect to the average constraints $I^*$ and $C^*$. 

\begin{equation} \label{fisher-i-beta}
\begin{split}
F_I(\nu) &= E \left[ \frac{\partial^2}{\partial \nu^2} \log P(I|\nu) \right]   \\
 &= E \left[\frac{\partial^2}{\partial \langle I^* \rangle^2} \log P(I|\langle I^* \rangle)   \left(\frac{\partial\langle I^* \rangle}{\partial \nu}\right)^2 \right]  
\end{split}
\end{equation}
The first term within the expectation operator is exactly the Fisher Information of $I$ with respect to $\langle I^* \rangle$, i.e., $F_I(\langle I^* \rangle)$, yielding
 \begin{equation} \label{fisher-i-beta2}
F_I(\nu)  = F_I(\langle I^* \rangle)\left(\frac{\partial\langle I^* \rangle}{\partial \nu}\right)^2
\end{equation} 
Analogously,
 \begin{equation} \label{fisher-c-beta2}
F_C(\nu)  = F_C(\langle C^* \rangle)\left(\frac{\partial\langle C^* \rangle}{\partial \nu}\right)^{2}
\end{equation}  

Thus, the Fisher Information of the joint variable $(I,C)$ with respect to the control parameter $\beta$ can be expressed via Fisher Information with respect to the conserved quantities $I^*$ and $C^*$: 
\[F_{I,C}(\nu) = F_I(\langle I^* \rangle)\frac{\partial\langle I^* \rangle^2}{\partial \nu} + F_C(\langle C^* \rangle)\left(\frac{\partial\langle C^* \rangle}{\partial \nu}\right)^{2}  \]

Using binomial random variables $(V$, $\langle I^* \rangle)$ and $(E$, $\langle C^* \rangle)$ in the expression for Fisher Information (\ref{fisher_binomial}), we can  produce our next result, directly expressing the Fisher Information of the MaxEnt distribution in terms of the conserved quantities  $I^*$ and $C^*$:
\begin{multline}
\label{fic2}
F_{I,C}(\nu) =\\
\frac{V}{\langle I^* \rangle (1 - \langle I^* \rangle)} \left(\frac{\partial\langle I^* \rangle}{\partial \nu}\right)^{2} + 
\frac{E}{\langle C^* \rangle (1 - \langle C^* \rangle)} \left(\frac{\partial\langle C^* \rangle}{\partial \nu}\right)^{2}  
\end{multline}

Importantly, the independence assumption, which allowed us to obtain  analytic expressions for the Fisher information (\ref{fisher-i-beta2})--(\ref{fic2}), reveals the mechanism for its divergence at criticality. It is known that the divergence of the Fisher information indicates a critical regime and pinpoints the critical threshold. In this instance, it shows that, when $0 < \langle I^* \rangle < 1$ or $0 < \langle C^* \rangle < 1$,  the phase transition should occur when the derivative of $\langle I^* \rangle$ or $\langle C^* \rangle$ with respect to the control parameter $\nu$ becomes infinite (i.e., has a vertical tangent). In other words, one of the implications of the independence assumption is that the mechanism behind criticality is explicitly linked to  non-differentiability of $\langle I^* \rangle$ or $\langle C^* \rangle$.

In order to explicitly distinguish between the Maxent distribution and the distributions observed from simulations, we  introduce $I_M, C_M$, the random variables distributed according to the MaxEnt solutions given in \eqref{binom_pic}; and $I_O,C_O$, the random variables associated to the observed probability distributions of our simulations (see Supplementary Materials S.1 and S.2 for details of numerically calculating the corresponding Fisher Information and thermodynamic efficiency of computation, respectively).

\section{\label{numres}Numerical results}
We constrain our analysis to the specific case of Watts-Strogatz random graphs~\cite{watts1998collective} across a broad range of simulated pathogens. Having considered the statistical mechanics of SIS processes under the assumption of independence, we will compare our analytical derivations with the results of numerical simulations of the steady state probability distributions. To reiterate, these probability distributions are obtained from computational simulations of stochastic discrete-time parallel-update SIS dynamics on Watts-Strogatz random graphs with 1000 nodes, parameters $k=6$ and $p=1$, varying  the probability of infection $\nu$ and holding the probability of recovery, $\delta=0.001$, constant. Probability distributions are observed for each of 30 graphs across 30 realisations over 100,000 timesteps per run for each value of $\nu$. For each run, we perform a logarithmic sweep of the parameter $\nu$ from $10^{-4}$ to $10^{-2}$ with 100 steps in this range. In order to obtain an appropriate representation of the stationary distributions, we do not record any data until the system has sufficient time to equilibrate (40,000 time steps). In addition to analysis of the steady state probability distributions, we study important thermodynamic quantities such as the entropy, Fisher Information and free entropy of the system.

\begin{figure}[!ht]
\begin{subfigure}{0.45\textwidth}
   \includegraphics[width=1\linewidth]{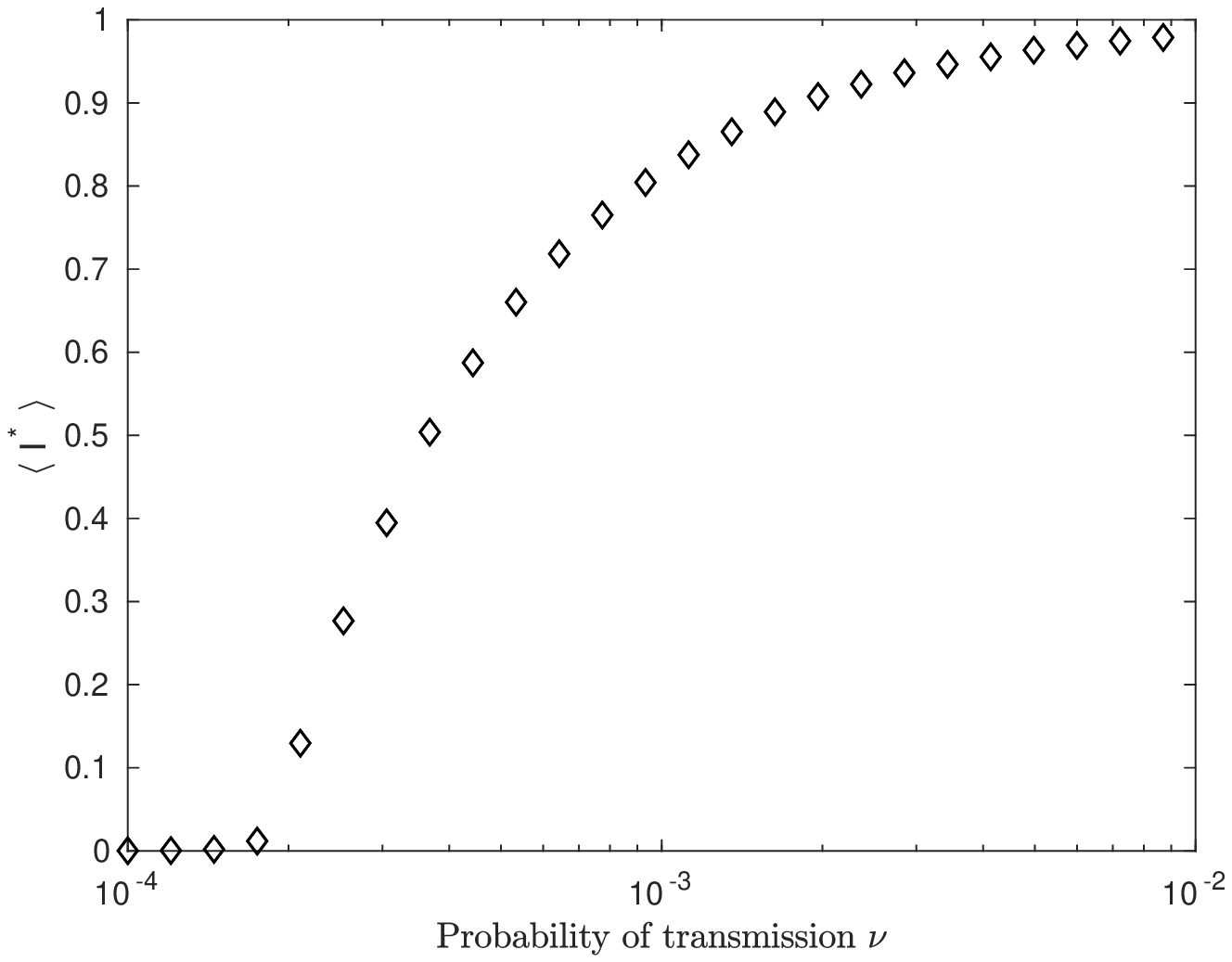}
   \caption{}
   \label{fig:I_constraint} 
\end{subfigure}

\begin{subfigure}{0.45\textwidth}
   \includegraphics[width=1\linewidth]{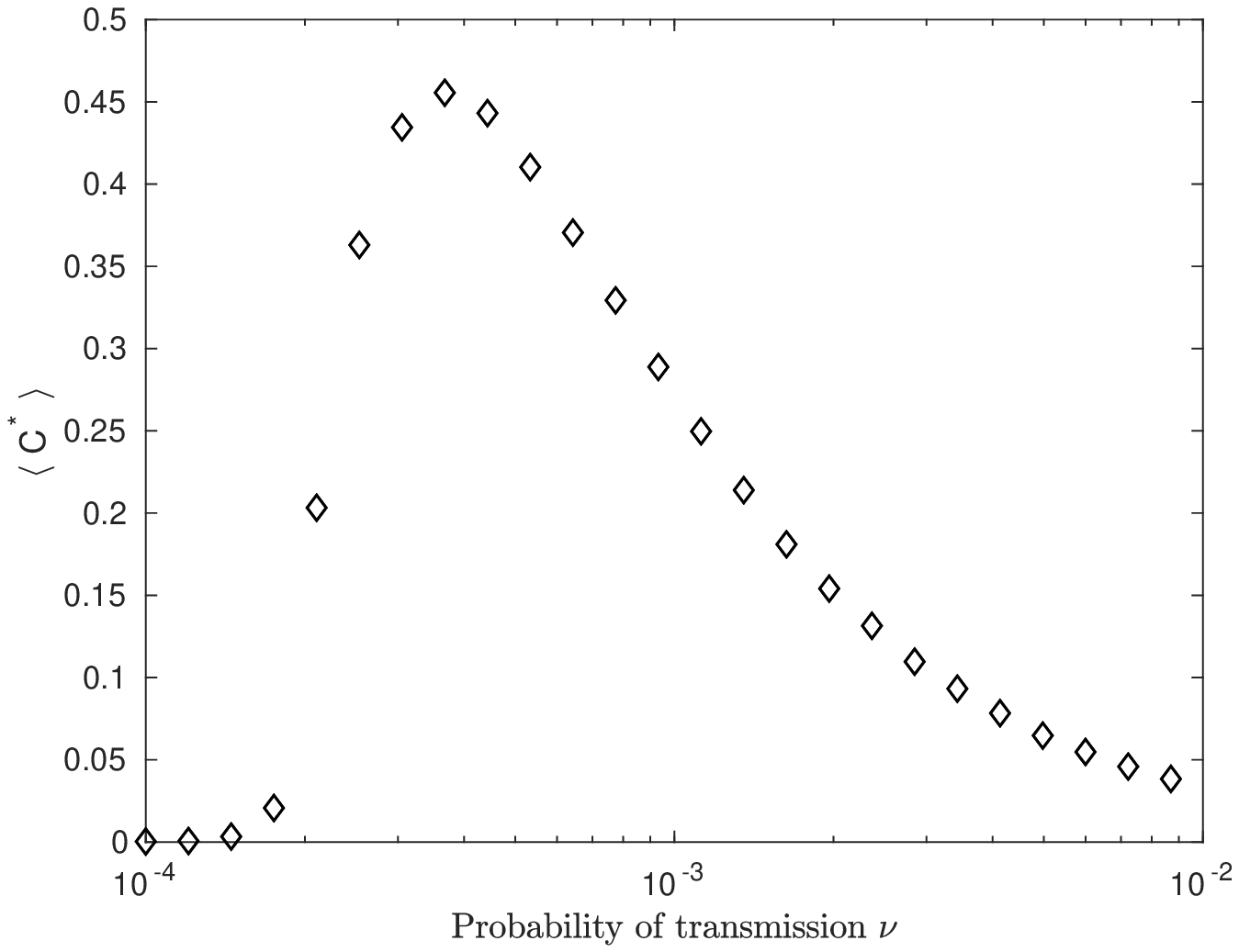}
   \caption{}
   \label{fig:C_constraint}
\end{subfigure}

\caption[Two constraints]{ The constraints $\langle I^* \rangle$ (a) and $\langle C^* \rangle$ (b) as functions of the probability of infection $\nu$.}
\label{fig:two_constraints}
\end{figure}

Firstly we study the constraints $\langle I^* \rangle$ and $\langle C^* \rangle$ as a function of the parameter $\nu$, the probability of infection, as shown in Fig.~\ref{fig:two_constraints}. These constraints are averaged over all simulations described above. $\langle I^* \rangle$ and $\langle C^* \rangle$ set the constraints  used in order to obtain the MaxEnt probability distribution.

\begin{figure}[!ht]
\begin{subfigure}[b]{0.45\textwidth}
   \includegraphics[width=1\linewidth]{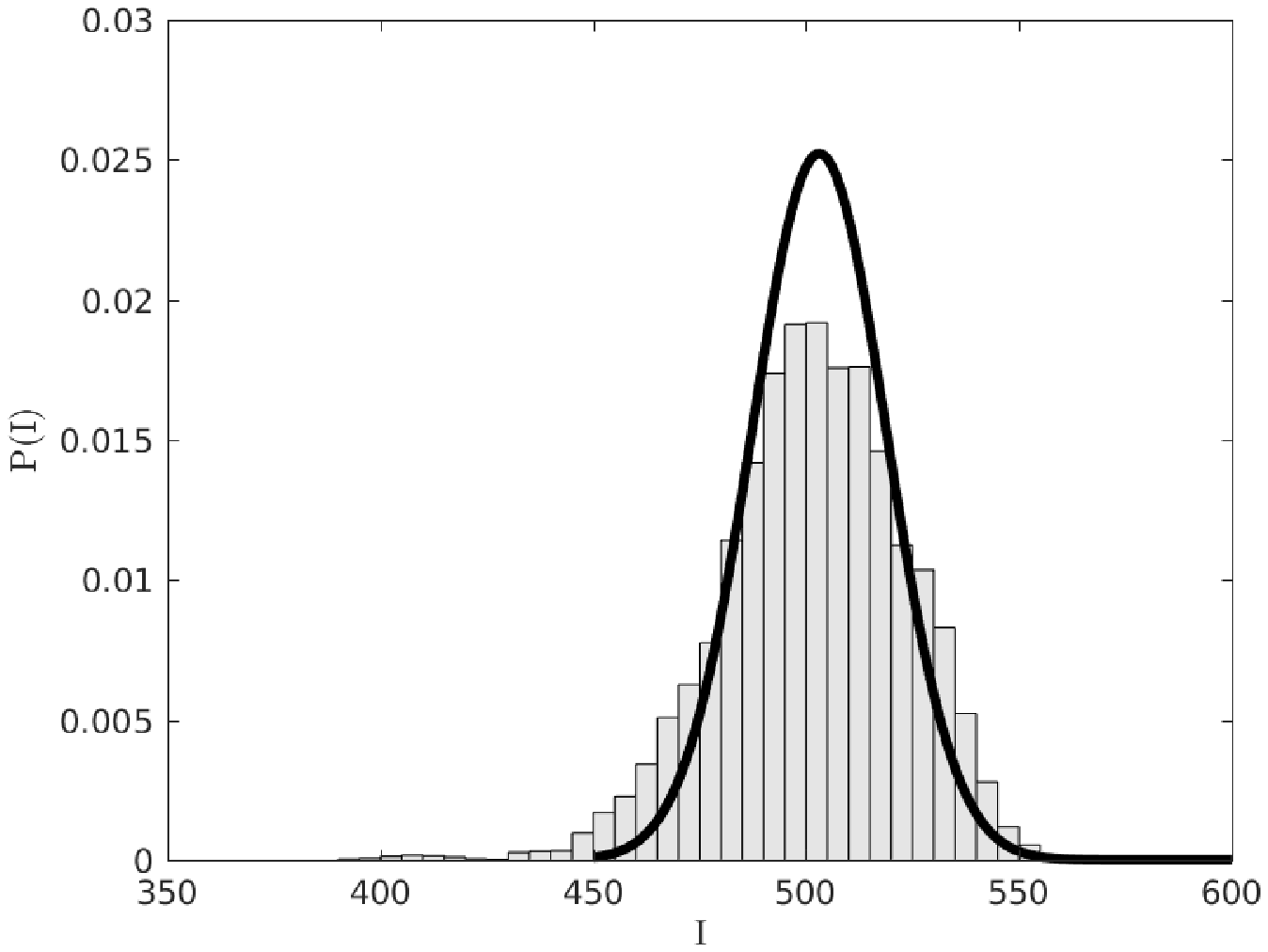}
   \caption{}
   \label{fig:I_prob_dist} 
\end{subfigure}

\begin{subfigure}[b]{0.45\textwidth}
   \includegraphics[width=1\linewidth]{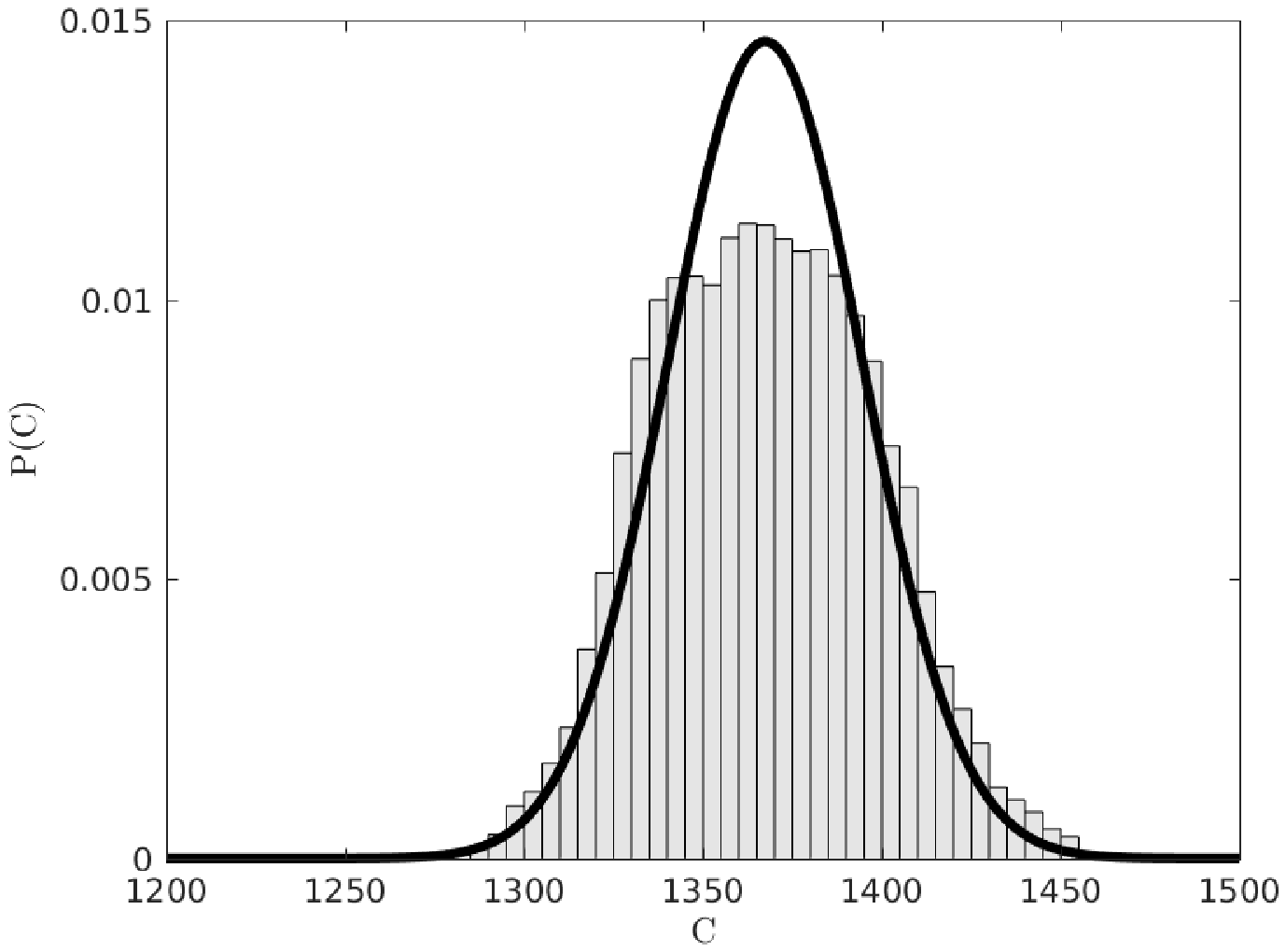}
   \caption{}
   \label{fig:C_prob_dist}
\end{subfigure}
\caption[Two_marginal_distributions]{ The observed simulation and MaxEnt probability density functions of $I$ (a) and $C$ (b). The histogram shows the empirically observed pdf with bin widths of 5, whereas the solid line gives the pdf of the MaxEnt solution for $\nu = 3.68 \times 10^{-4}$ and $\delta = 0.001$.}
\label{fig:distributions}
\end{figure}

\begin{figure}[!ht]
\begin{subfigure}[b]{0.45\textwidth}
   \includegraphics[width=1\linewidth]{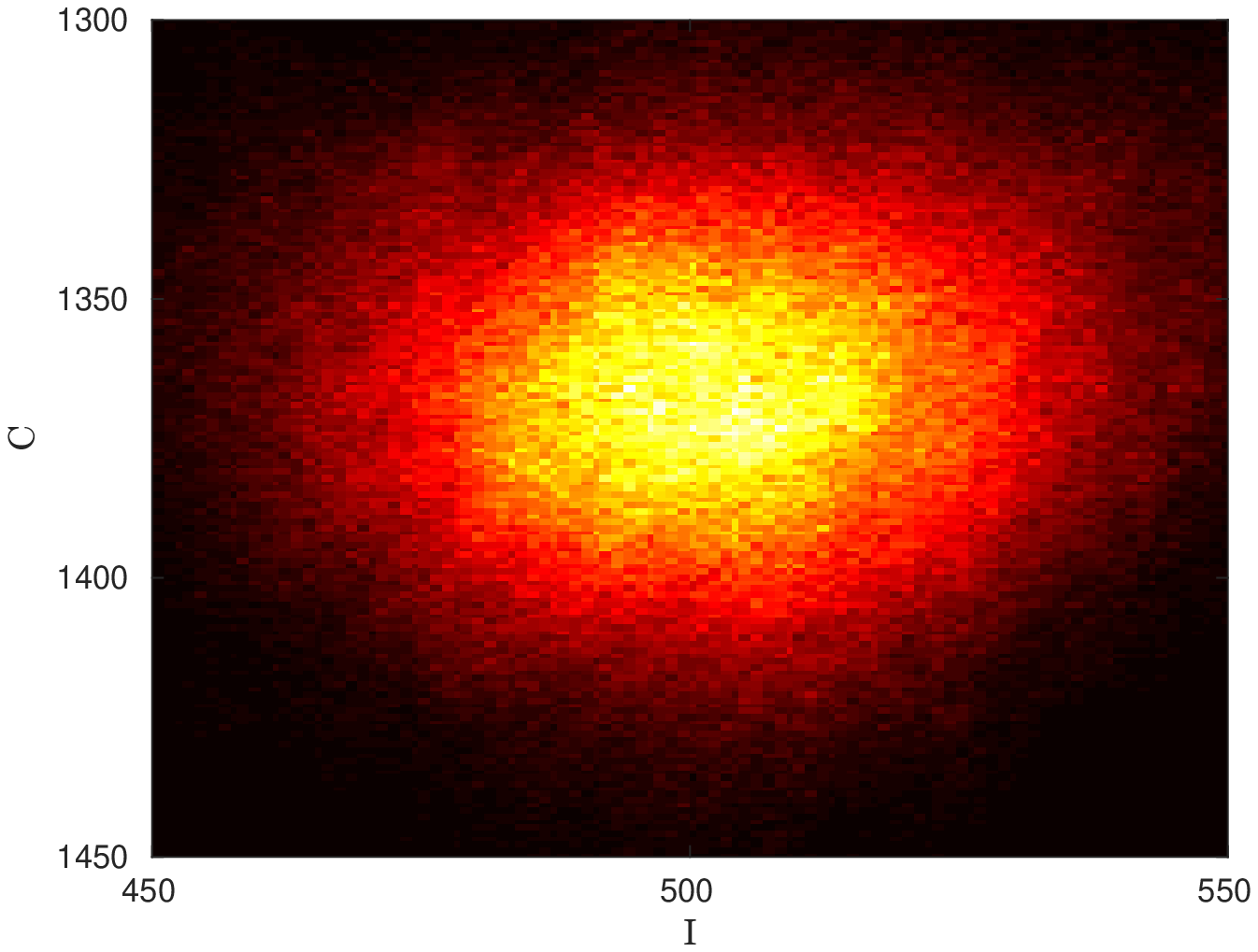}
   \caption{}
   \label{fig:IC_distribution} 
\end{subfigure}

\begin{subfigure}[b]{0.45\textwidth}
   \includegraphics[width=1\linewidth]{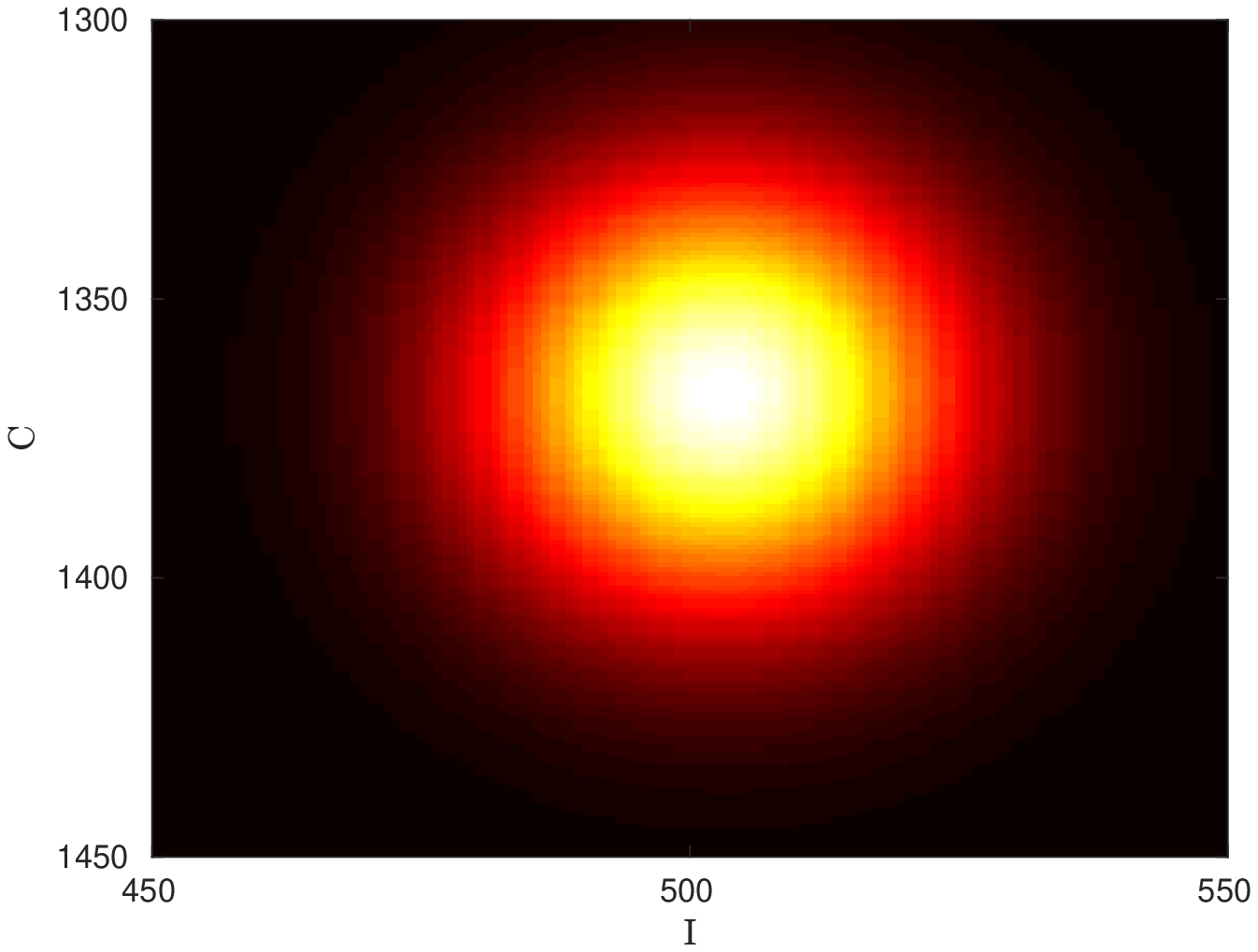}
   \caption{}
   \label{fig:IC_distribution_maxent}
\end{subfigure}

\caption[Joint distributions]{ The observed (a) and the MaxEnt (b) probability density functions of the joint random variable $(I,C)$. The probability distributions are generated from simulations with $\nu = 3.68 \times 10^{-4}$ and $\delta = 1 \times 10^{-3}$.}
\label{fig:joint_distributions}
\end{figure}

Secondly, we investigate the two types of marginal and joint probability distributions, shown in Figures~\ref{fig:distributions} and~\ref{fig:joint_distributions} respectively. Each of the figures shows both the distribution obtained empirically from computer simulations and the distribution obtained from the MaxEnt solution~\eqref{binom_pic}, given the average quantities. Specifically, the empirical distributions  allowed us to estimate the average quantities $\langle I \rangle$ and $\langle C \rangle$, shown earlier  in Figures~\ref{fig:I_constraint} and \ref{fig:C_constraint}, used in deriving the MaxEnt solutions. These results are obtained from simulations with $\nu = 3.68 \times 10^{-4}$ and $\delta = 0.001$, where $\nu$ is the per time step probability of infection and $\delta$ is the per time step probability of recovery.
Fig. \ref{fig:distributions} compares the marginal distributions of $I$ and $C$ observed from simulations to the maximum entropy distributions obtained only from the observations of the average values $\langle I \rangle$ and $\langle C \rangle$ on a particular Watts-Strogatz graph. Secondly, Figures~\ref{fig:IC_distribution} and \ref{fig:IC_distribution_maxent} demonstrate the differences between the experimentally observed distribution of $(I,C)$ and the distributions of $(I,C)$ according to the Maximum Entropy principle.

\subsection{Fisher Information}
As mentioned earlier, it is well known that the Fisher Information diverges at a phase transition~\cite{prokopenko2011relating}. We now  compare the Fisher Information for both the MaxEnt and observed distributions. This comparison is carried out with the change of variables given by equation~(S.1) with $u_{min}=-4$ and $\Delta u = 4.65 \times 10^{-2}$ for 100 values of $u_i = u_{min} + i \Delta u$, using (S.7) to calculate the Fisher Information of the observed distributions.  
For each value of $\nu$, a single probability distribution is obtained for each graph using multiple runs, and then the Fisher Information is calculated, as outlined in section \ref{numres}, by averaging over each of the graphs.

\begin{figure}[!ht]
\begin{subfigure}[b]{0.4\textwidth}
   \includegraphics[width=1\linewidth]{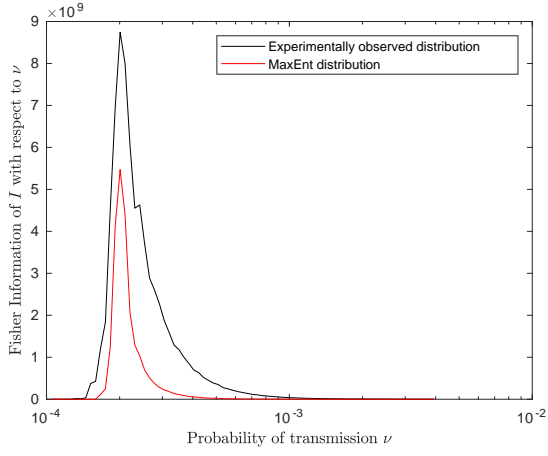}
   \caption{}
   \label{fig:fisher_I}
\end{subfigure}
\begin{subfigure}[b]{0.4\textwidth}
   \includegraphics[width=1\linewidth]{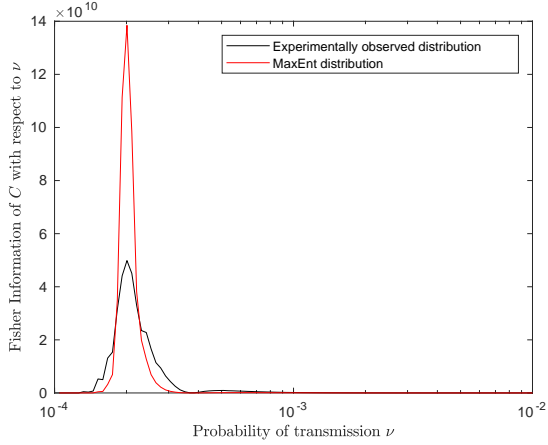}
   \caption{}
   \label{fig:fisher_C}
\end{subfigure}
\begin{subfigure}[b]{0.4\textwidth}
   \includegraphics[width=1\linewidth]{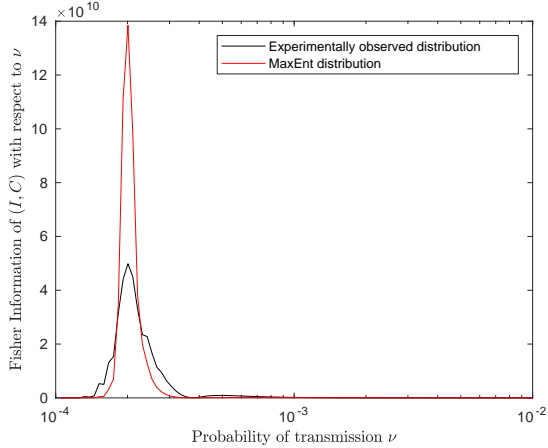}
   \caption{}
   \label{fig:fisher_IC}
\end{subfigure}
\caption{The Fisher Information of (a) $I$, (b) $C$ and (c) the joint variable $(I,C)$ as a function of $\nu$, the probability of infection per time step. The value of the Fisher Information is an average of values calculated for individual graphs. The peak of each of these plots occurs at $\nu = 2.01 \times {10^{-4}}$, corresponding to $R_0 = 1.004$. }
\label{fig:fisher_critical}
\end{figure}

As shown in Fig.~\ref{fig:fisher_critical}, for $\delta = 1 \times 10^{-3}$, the Fisher Information of each of the random variables peaks around $2.01 \times {10^{-4}}$. For this system, this corresponds to $R_0 = 1.004$ (see equation (S.16)). Due to uncertainty in the location of this maximum, $\nu \in ( 1.9179\times 10^{-4} , 2.1049\times 10^{-4} )$, there is corresponding uncertainty in the value of $R_0$. In this case, $R_0$ is in the interval $( 0.9657 , 1.0435 )$.  Importantly, the approach identifies the critical threshold for the epidemic phase transition.


As pointed out in section \ref{therm_eff_comp}, under a quasistatic protocol changing the control parameter $\nu$, the Fisher Information may also be interpreted as the generalised work, $W_{gen}$~\eqref{eq:fisher-work-curvature}.

\subsection{Entropy and free entropy}
\begin{figure}[!ht]
\begin{subfigure}[b]{0.4\textwidth}
   \includegraphics[width=1\linewidth]{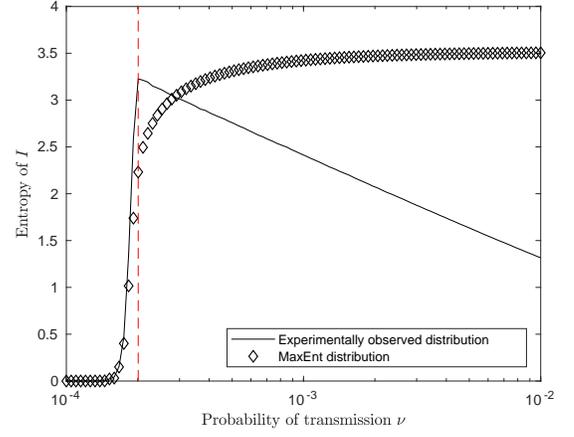}
   \caption{}
   \label{fig:entropy_I}
\end{subfigure}
\begin{subfigure}[b]{0.4\textwidth}
   \includegraphics[width=1\linewidth]{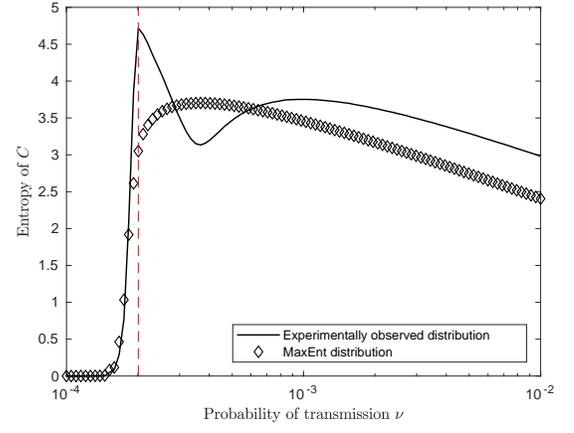}
   \caption{}
   \label{fig:entropy_C}
\end{subfigure}
\begin{subfigure}[b]{0.4\textwidth}
   \includegraphics[width=1\linewidth]{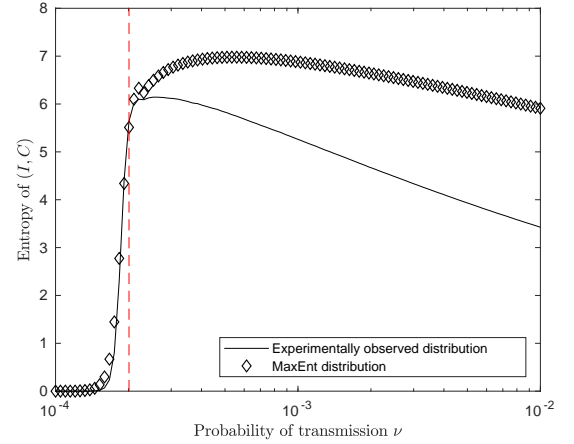}
   \caption{}
   \label{fig:entropy_IC}
\end{subfigure}
\caption{The Shannon entropy of the random variables (a) $I$, (b) $C$ and (c) $(I,C)$ as a function of $\nu$, the probability of transmission. The dotted line indicates the peak of the Fisher Information.}
\end{figure}

We now turn our attention to other thermodynamic characteristics, beginning with the configuration entropy~\eqref{Shan_ent}, and again comparing the observed and the MaxEnt distributions across a  range of values of $\nu$. Fig.~\ref{fig:entropy_I} shows that, after a phase transition, there is decreasing log-linear relationship between the entropy of $I$ and $\nu$ for the distributions obtained from the simulations. In contrast, the entropy of the MaxEnt distributions does not show this dependency, resulting in an asymptotic overestimate.

\begin{figure*}[!ht]
\includegraphics[width = 0.9\textwidth]{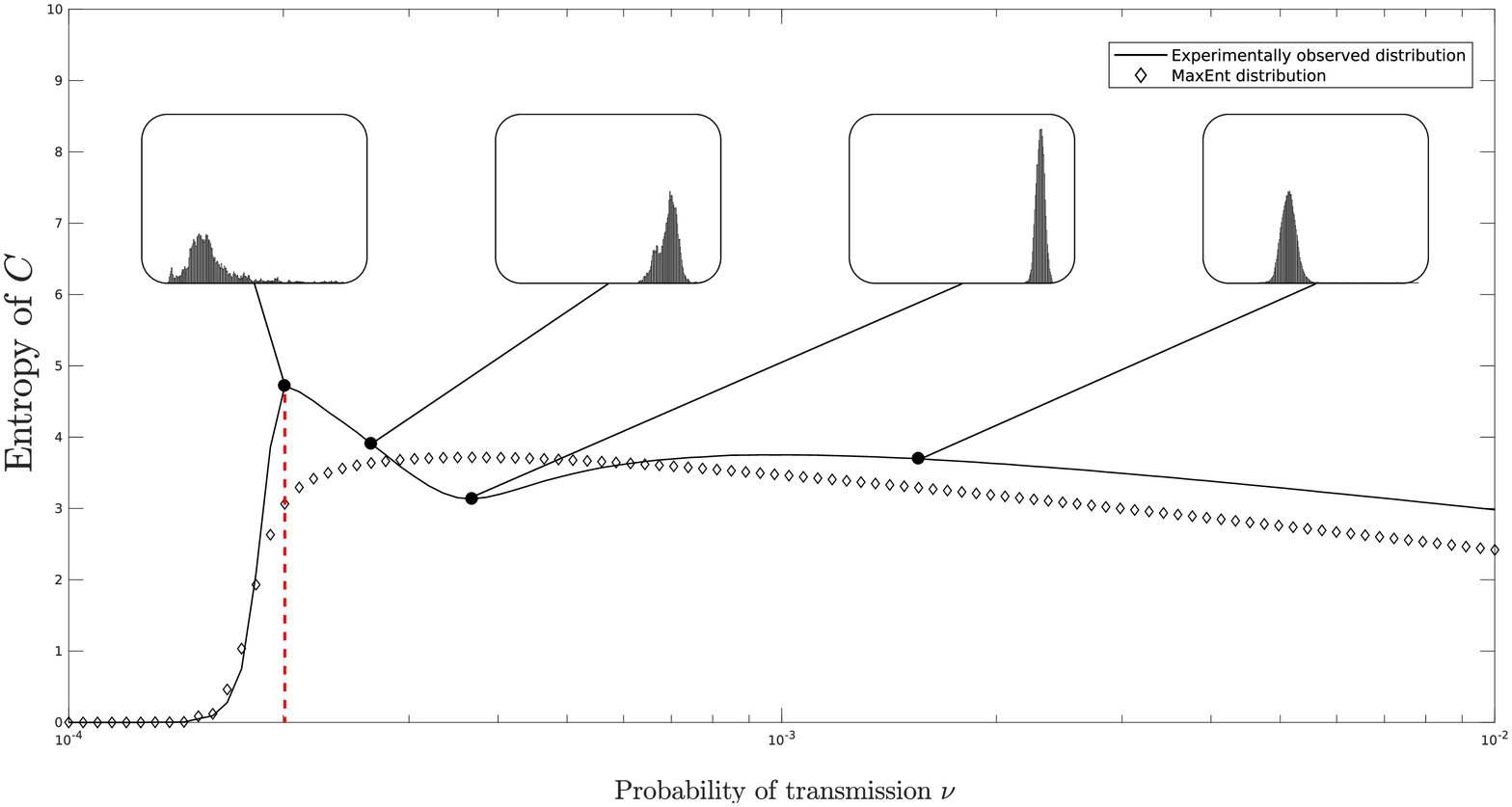}
\caption{\label{detailed_entropy_C} The entropy of the random variable $C$ as a function of the parameter $\nu$ for distributions obtained from simulations and the MaxEnt distributions. The dotted line indicates the value of $\nu$ for which the Fisher Information peaks, i.e. the critical value of $\nu$. Shown inset for comparison are the observed probability distributions for several values of $\nu$.}
\end{figure*}

We also note that although the entropy of the MaxEnt solution for $C$ captures the overall trend of the observed entropy, as shown in Fig.~\ref{fig:entropy_C}, it does not precisely capture the qualitative behaviour of the entropy of $C$, ``smoothing'' the drop in entropy immediately following the phase transition. Fig.~\ref{detailed_entropy_C} shows the distributions of $C$ inset for multiple values of $\nu$.  Furthermore, we note a good general agreement for the entropy of the joint variable, illustrated by Fig.~\ref{fig:entropy_IC}.

These discrepancies and similarities clarify the impact of the independence assumption on the thermodynamics of the epidemics: despite disagreements in a ``super-critical'' phase, the phase transition itself has been identified equally well by both the MaxEnt distributions and the distributions obtained from the simulations.

\subsection{Thermodynamic efficiency of computation}

\begin{figure}[!ht]
\includegraphics[width=1\linewidth]{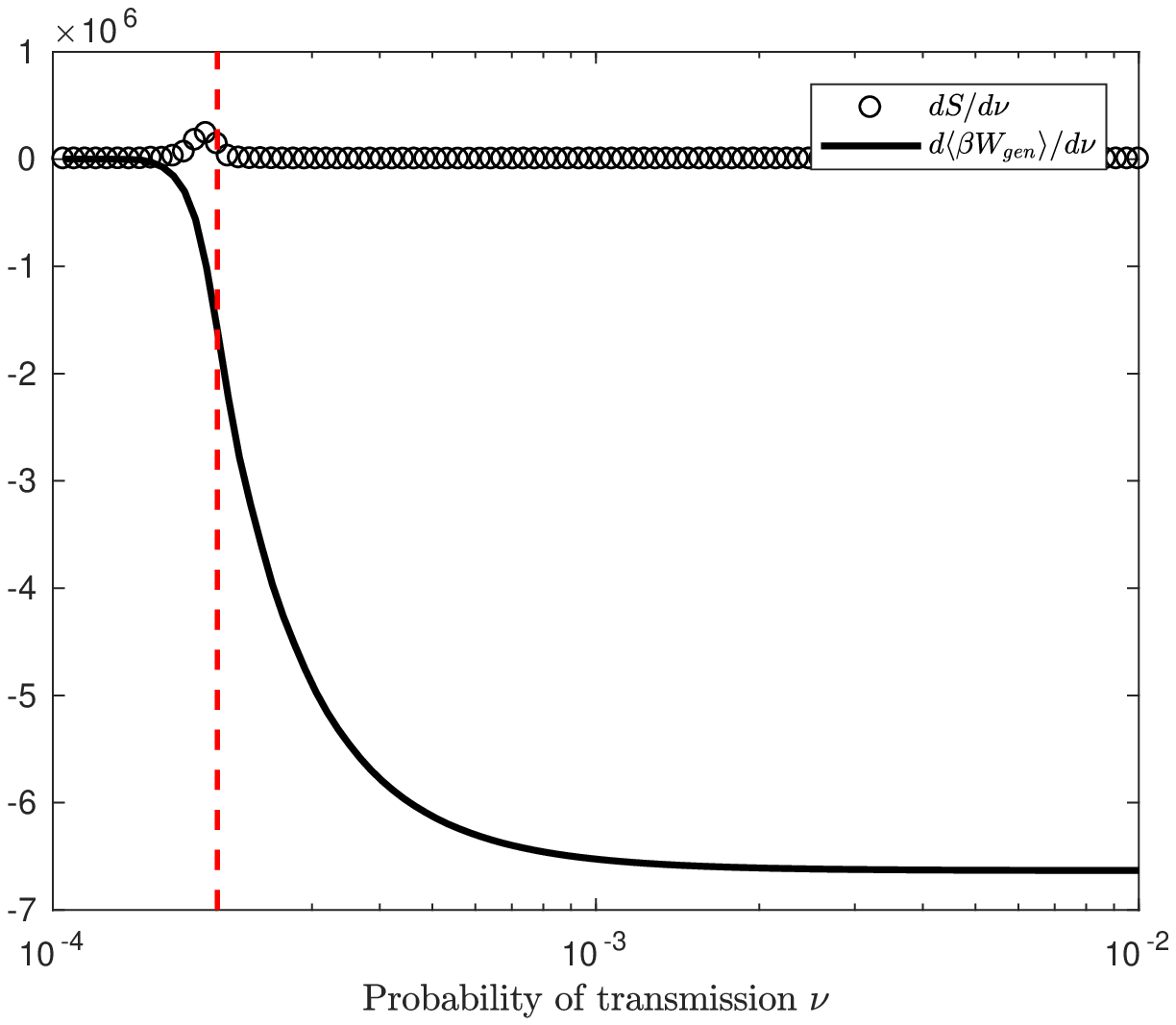}
\caption{\label{fig:therm_efficiency_parts}Rates of change of work and configuration entropy with respect to $\nu$, the probability of of transmission, for observed distributions. The dotted line indicates the critical point at which the Fisher Information peaks.}
\end{figure}

\begin{figure}[!ht]
\includegraphics[width = 0.4\textwidth]{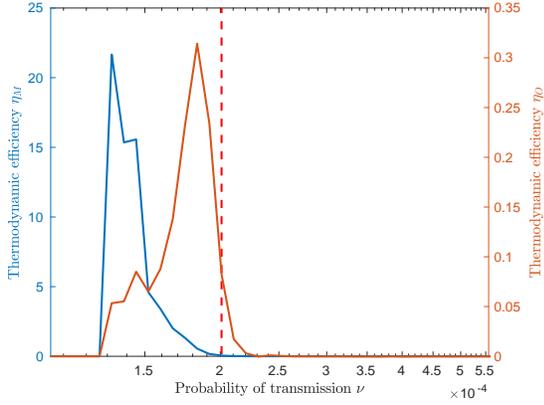}
\caption{\label{fig:therm_efficiency_two_plots}Thermodynamic efficiency of computation  against the probability of infection $\nu$: in red for observed distributions, in blue for the MaxEnt distributions.
The  dotted line indicates the critical point at which the Fisher Information peaks.}
\end{figure}

\begin{figure}[!ht]
\includegraphics[width = 0.4\textwidth]{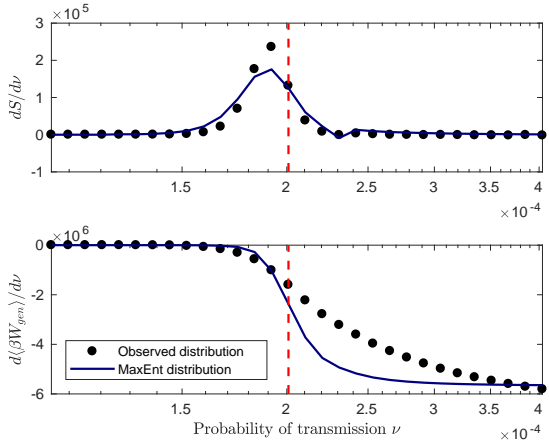}
\caption{\label{fig:therm_efficiency_parts_comparison}Rates of change of the configuration entropy~(top) and generalised work~(bottom). The dotted line indicates the critical value of $\nu$.}
\end{figure}

Fig.~\ref{fig:therm_efficiency_parts} illustrates the two key components of the thermodynamic efficiency of computation. We observe three qualitatively distinct regions in the space of $\nu$, which can be associated with sub-critical, critical and super-critical regions. Most strikingly, we note that in both the sub-critical and super-critical regions, the thermodynamic efficiency is very low, while around criticality, the system is most efficient, as shown in Figure~\ref{fig:therm_efficiency_two_plots}. The peak of the thermodynamic efficiency does not concur precisely with the peak of the Fisher Information, due to finite-size effects, and the discrepancy is higher for the estimates based on the MaxEnt method, rather than those based on the observed distributions. The specific reasons for this discrepancy can be seen in Figure~\ref{fig:therm_efficiency_parts_comparison}, which contrasts both components in more detail, showing that the reduction in uncertainty (the numerator) estimated by the MaxEnt method starts to diverge earlier, while the rate of work (the denominator) estimated by the MaxEnt method starts to diverge later, than the corresponding estimates based on observed distributions. As a result, the ratio $\eta_M$ ``suffers'' from the finite-size effects more than $\eta_O$.

In summary, at criticality, where a higher disorder is generated, there is relatively more work extracted out of the system. In an epidemiological context, considering an intervention process that reduces the transmission probability from the super-critical phase towards the sub-critical phase, expending the work, the thermodynamic efficiency would tend to increase as the critical point is passed. On the other hand, we can consider the efficiency of the contagion itself, as a biological phenomenon, i.e., a pathogen emergence process that increases the transmission probability from the sub-critical phase towards the super-critical phase. In this case, from the pathogen ``perspective'', the thermodynamic efficiency  would tend to peak on the approach towards the critical point.

\begin{figure}[!ht]
\includegraphics[width = 0.4\textwidth]{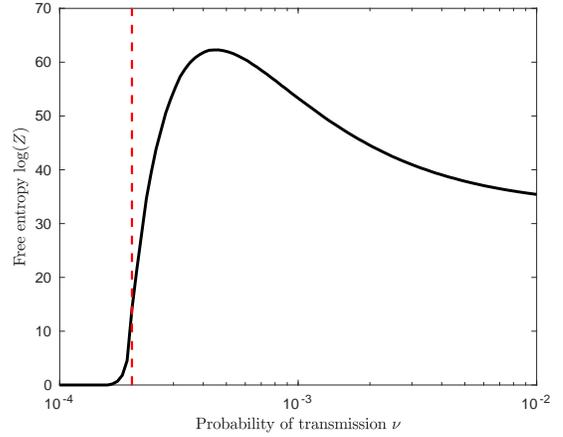}
\caption{The free entropy of the MaxEnt distributions. The dotted line indicates the critical value of $\nu$.}
\label{Ln_Z}
\end{figure}

Finally, Fig.~\ref{Ln_Z} shows the free entropy of the system, $\log(Z)$, which is proportional to the free energy. Again, we observe a clear critical regime separating the ``sub-critical'', non-epidemic, phase from the ``super-critical'', epidemic, phase.

\section{Discussion and future work}
In this paper we considered epidemics as thermodynamic phenomena, modelling SIS dynamics on a contact network statistical-mechanically. Applying the Maximum Entropy principle, under certain assumptions, allowed us to derive closed form solutions for specific cases and interpret key epidemic variables in a statistical mechanical setting.  Specifically, the reproductive ratio $R_0$ of a SIS model was related to the inverse temperature of a Gibbs distribution resulting from entropy maximisation, applied to the initial state of an outbreak. 

Using the model, we evaluated the Fisher Information, configuration and free entropy, as well as the thermodynamic efficiency of contagion (considered as a computational process), in an epidemiological context. This allowed us to identify critical regimes and distinct phases of epidemic processes.  Analytical derivations of Maximum Entropy modelling of SIS process were contrasted with results of the simulated dynamics on Watts-Strogatz random graphs, confirming the critical thresholds, while assessing the impact of our simplifying assumptions. Importantly, the analytical derivations highlighted the mechanism for the emergence of critical regime --- via non-differentiability of key variables $\langle I \rangle$ and $\langle C \rangle$ --- manifesting itself in the divergence of the Fisher Information. 

The statistical-mechanical perspective taken in this work placed  epidemic processes within a broad class of distributed  processes. This allowed us to define thermodynamic efficiency of contagions which was shown to peak at criticality. Interestingly, this can be interpreted as both (i) the efficiency of an intervention process that expends the work needed to reduce the transmission probability, or (ii) the efficiency of the pathogen emergence that extracts the work by increasing the transmission probability. These processes explore the parameter-space in opposing directions at the expense/extraction of (thermodynamic) work, but despite the opposite directions  the maximal thermodynamic efficiency is attained at the same critical point. Furthermore, the concept of thermodynamic efficiency enables comparative analysis of various interventions and pathogen emergence paths.

We presented both numerical and analytical techniques based on the MaxEnt method. The numerical technique does not need the independence assumption between $I$ and $C$, and attains a high degree of accuracy. However, it requires an explicit combinatorial calculation of the macrostates which is prohibitive for large networks without further simplifying assumptions. The analytical technique allowed us to derive important general thermodynamic properties of the contagion phenomena, but relies on the independence assumption and is less accurate in the super-critical phase. We believe that both presented techniques provide useful arguments for the utility of the MaxEnt method and serve their distinct purposes.

We would like to reiterate that in practical scenarios, when simulation results are not available and the exact topology of the underlying interaction network is not known, one may still rely on the MaxEnt solutions derived under the constraints on $\langle I \rangle$ and $\langle C \rangle$ which are obtainable from the real-world observations of the epidemic dynamics. If, on the other hand, the underlying interaction network belongs to a specific type, e.g., Watts-Strogatz random graph, one may go one step further and, under the simplifying independence assumption, obtain analytic solutions and estimate their parameters.

Determining the precise range of applicability of the independence assumption, and the extent of the resultant approximations, remains the subject of future work.  In general, this assumption is reasonable in networks with a significant number of random connections, and so would be applicable to many real-world networks in which epidemics take place, e.g., epidemics in urbanised societies \cite{pastor2015epidemic,neiderud2015}.

We believe that the presented approach can be extended to other contact network topologies and contact processes.  In addition, the constraints used during the entropy maximisation can be effectively generated by large-scale simulations of epidemics based on demographic datasets and real-world epidemic dynamics.

\section*{Data Accessibility}
Code used to generate the data used in this paper can be found at https://github.com/NathanHarding/Therm-eff-cont.

\section*{Authors' contributions}
All authors contributed to conception, analysis, interpretation and  drafting. N.H. performed computational analysis. R.N. carried out the work described in section \ref{complete_graph_ex}. All authors gave final approval for publication.

\section*{Competing interests}
We have no competing interests.

\section{Acknowledgements}
This research was supported by the Sydney Informatics Hub, through the use of HPC services funded by the University of Sydney. The authors would like to thank Emanuele Crosato for numerous helpful discussions on the intricacies of thermodynamic efficiency.

\section*{Funding}
M.P. and N.H. were supported through the Australian Research Council grant DP160102742. Additionally, N.H. was supported by an Australian Government Research Training Program (RTP) Scholarship.

\appendix
\section{\label{sec:fisher_est}Estimation of Fisher Information}
To trace the Fisher Information over a logarithmic scale, we substitute 
\begin{equation}\label{u_sub}
u = \log_{10}(\nu)
\end{equation}
so that
\begin{equation}\label{u_deriv}
\frac{du}{d\nu} = \frac{1}{\nu \log{10}} = \frac{10^{-u}}{\log(10)}.
\end{equation}
This substitution is chosen for smoothness and resilience to noise as the difference between distributions becomes significant, the effect of noise on the calculation of the Fisher Information is diminished.

The Fisher Information of $I$, $C$ and the joint random variable $(I,C)$ with respect to $\nu$ can be expressed using the reparametrisation of the Fisher Information~\eqref{Fisher_reparam} as follows:
\begin{equation}\label{Fisher_ME_num}
F_{I_{M}}(\nu) = F_{I_{M}}(\langle I^* \rangle) \left( \frac{d\langle I^* \rangle}{d u} \right)^2 \left( \frac{du}{d \nu} \right)^2
\end{equation}

\begin{equation}\label{Fisher_Obs_num}
F_{I_{O}}(\nu) = F_{I_{O}}(u) \left( \frac{du}{d \nu} \right)^2
\end{equation}
Substituting equations~\eqref{u_deriv} and \eqref{fisher_binomial} with $n=V$ and $q = \langle I^* \rangle$ into equations~\eqref{Fisher_ME_num} and \eqref{Fisher_Obs_num} yields
\begin{equation}\label{Fisher_ME_num_2}
F_{I_{M}}(\nu) = \frac{V}{\langle I^* \rangle (1 - \langle I^* \rangle)} \left( \frac{d\langle I^* \rangle}{d u} \right)^2 \left( \frac{1}{\nu \log{10}} \right)^2
\end{equation}
and
\begin{equation}\label{Fisher_Obs_num_2}
F_{I_{O}}(\nu) = F_{I_{O}}(u) \left( \frac{1}{\nu \log{10}} \right)^2
\end{equation}

When there is no closed form expression for the Fisher Information, it can be estimated numerically using suitable discretisations over system states $x_1, \ldots, x_n$. The derivative with respect to $\lambda$ is calculated using a finite difference method with step length $\Delta \lambda$. In this paper we will use the backwards finite difference method to approximate the Fisher Information, denoted $\hat{F}_X(\lambda)$: 
\begin{multline}\label{fisher_numerical}
\hat{F}_X(\lambda) = \\
\sum_{i=1}^N p(x_i;\lambda) \left( \frac{\log \left( p(x_i;\lambda)\right)-\log \left( p(x_i;\lambda-\Delta \lambda)\right)}{\Delta \lambda} \right)^2
\end{multline}

\section{Thermodynamic efficiency of computation $\eta$}
\label{sec:TEC2}
To determine the thermodynamic efficiency of computation $\eta$, as defined in equation~\eqref{TEC}, we firstly identify the zero-response point\eqref{0rp}. For our system it is simply zero, as there is no work expended in changing the transmission probability near zero, that is, $\nu^*=0$. Hence, using \eqref{work_deriv}, we obtain:
\begin{equation}
\frac{d\langle \beta W_{gen}\rangle}{d\nu} = -\int_0^\nu F_{I,C}(\nu') d\nu'
\end{equation}
Again, we substitute $u = \log_{10}(\nu)$, resulting in:
\begin{equation}\label{eq:numerical_ds_dnu}
\frac{dS}{d \nu} = \frac{dS}{du} \frac{du}{d\nu} =\frac{dS}{du} \frac{10^{-u}}{\log{10}}
\end{equation}
and
\begin{multline}
\frac{d \langle \beta W_{gen}\rangle}{d\nu} = - \int_{u(0)}^{u(\nu)} F_{I,C}(\nu'(u')) \frac{d\nu'}{du'}du' \\
 = - \int_{u(0)}^{u(\nu)} F_{I,C}(\nu'(u')) 10^{u'} \log(10)du'
\end{multline}
These quantities are both calculated numerically for 100 values of $u$ between $-4 \leq u \leq -2$. With 
$\frac{dS}{d \nu}$ calculated numerically using a backwards difference method, and ${d \langle \beta W_{gen}\rangle/d\nu}$ calculated using a cumulative trapezoidal numerical integration.

The entropy in proximity of the critical point was approximated using a nonlinear least-squares fit to a logistic growth curve. The fitted values for the entropy were used exclusively for the calculation of the numerical derivative $dS/du$ in the calculation of $\eta$ in the interval $\nu = [1.26\times 10^{-4},2.21\times 10^{-4}] $.

We approximate the entropy of $(I,C)$ as a logistic growth curve for $\nu = 1 \times 10^{-4}$ to $2.9 \times 10^{-4}$. Explicitly, we fit
\begin{equation}\label{entropy_fit}
S_{fit} = \frac{L}{1+e^{-k(\nu-\nu_0)}}.
\end{equation}

\begin{figure}[!th]
\includegraphics[width = 0.9\linewidth]{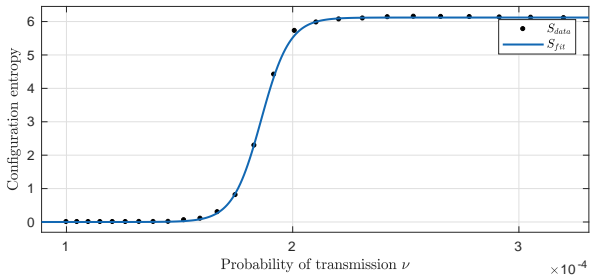}
\caption{Comparison of configuration entropy of the joint random variable $(I,C)$ and the best fit $S_{fit}$.}
\label{fig:fit}
\end{figure}

Using the MATLAB curve-fitting tool, we obtain a curve of best fit with $L = 6.12$, $k =1.67 \times 10^{5}$ and $\nu_0 = 1.855 \times 10^{-4}$, see Fig.~\ref{fig:fit}. For the data on $\nu = 1 \times 10^{-4}$ to $2.9 \times 10^{-4}$, this curve has an R-squared value of $0.99$.

\section{Relating $R_0$ to probability of infection}
We show that, for the system described in this study, the reproductive ratio $R_0$ can be expressed analytically as a function of $\nu$, $\delta$ and $k$, the average degree of the graph. 
Let $Y$ be the event when some infected individual $x$, before recovering from the infection, infects a neighbour $y$. Let  us also denote   the degree of $x$ as $k_x$. Then 
\begin{equation}
R_0 = E[k_x E[Y]].
\end{equation}
In the case of the Watts-Strogatz graph, the average degree within the population is equal to k, and so
\begin{equation}
R_0 = k E[Y]
\end{equation}
In this case, a closed form of $P(Y)$ is given by
\begin{multline}
 P(Y) = \nu+\nu(1-\nu)(1-\delta) +\nu (1-\nu)^2 (1-\delta)^2 + \ldots \\
= \nu\sum_{j = 0}^\infty (1-\nu)^j(1-\delta)^j  
\end{multline}
where the $j^{th}$ term of the sum represents the probability of $j$ consecutive unsuccessful infections of $y$ by $x$,  and $j$ unsuccessful recoveries of $x$, followed by one successful infection. Since $\nu$ is in the interval $[0,1]$ and $\delta$ is in the interval $[0,1]$, this geometric series converges to 
\begin{figure}[!ht]
\includegraphics[width = 0.9\linewidth]{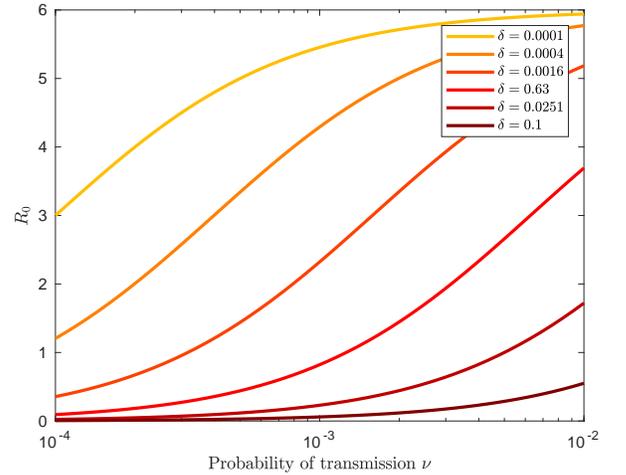}
\caption{The reproductive ratio $R_0$ as a function of $\nu$ varying $\delta$}
\end{figure}
\begin{equation}
\frac{\nu}{1-(1-\nu)(1-\delta)} = \frac{\nu}{\nu+\delta-\nu\delta}
\end{equation}
Thus, 
\begin{equation}\label{R_0_formula}
R_0 = \frac{k\nu}{\nu+\delta-\nu\delta}
\end{equation}

\bibliography{Max_ent_SIS}
\end{document}